\documentclass[amssymb,aps,prl,superscriptaddress,noeprint,reprint,twocolumn]{revtex4-1}

\usepackage{graphicx}
\usepackage{mathtools}
\usepackage{tikz}
\usepackage{siunitx}
\usepackage{xcolor}
\usepackage[normalem]{ulem}
\usepackage{hyperref}

\usepackage{xcolor}
\hypersetup{
    colorlinks,
    linkcolor={red!50!black},
    citecolor={blue!50!black},
    urlcolor={blue!80!black}
}

\usepackage{color}
\definecolor{blue}{rgb}{0.00,0.00,0.95}

\usepackage{pgffor}
\usepackage{pdfpages} 
\makeatletter
\AtBeginDocument{\let\LS@rot\@undefined}
\makeatother

\begin{document}

\title{Inferring the dynamics of underdamped stochastic systems}

\author{David B. Br\"uckner}
\thanks{These authors contributed equally.}
\affiliation{Arnold Sommerfeld Center for Theoretical Physics and Center for NanoScience, Department of Physics, Ludwig-Maximilian-University Munich, Theresienstr. 37, D-80333 Munich, Germany}
\author{Pierre Ronceray}
\thanks{These authors contributed equally.}
\affiliation{Center for the Physics of Biological Function, Princeton University, Princeton, NJ 08544, USA}
\author{Chase P. Broedersz}
\email[]{c.broedersz@lmu.de}
\affiliation{Arnold Sommerfeld Center for Theoretical Physics and Center for NanoScience, Department of Physics, Ludwig-Maximilian-University Munich, Theresienstr. 37, D-80333 Munich, Germany}
\affiliation{Department of Physics and Astronomy, Vrije Universiteit Amsterdam, 1081 HV Amsterdam, The Netherlands}

\begin{abstract}
  Many complex systems, ranging from migrating cells to animal groups,
  exhibit stochastic dynamics described by the underdamped Langevin
  equation. Inferring such an equation of motion from experimental
  data can provide profound insight into the physical laws governing
  the system. Here, we derive a principled framework to infer the
  dynamics of underdamped stochastic systems from realistic
  experimental trajectories, sampled at discrete times and
  subject to measurement errors. This framework yields an operational
  method, Underdamped Langevin Inference (ULI), which performs well on
  experimental trajectories of single migrating cells and in complex
  high-dimensional systems, including flocks with Viscek-like alignment
  interactions. Our method is robust to experimental measurement errors, and includes a
  self-consistent estimate of the inference error.
\end{abstract}

\maketitle

Across the scientific disciplines, data-driven methods are used to
unravel the dynamics of complex systems. These approaches often take
the form of inverse problems, aiming to infer the underlying
governing equation of motion from observed trajectories. This problem
is well understood for deterministic
systems~\cite{Crutchfield1987,Daniels2015,Brunton2015}. For a broad
variety of physical systems, however, a deterministic description is
insufficient: fast, unobserved degrees of freedom act as an effective
dynamical noise on the observable quantities.
Such systems are described by Langevin dynamics, and inferring their equation of motion is notoriously harder: one must then disentangle the stochastic from the deterministic contributions, both of which contribute to shape the trajectory. In molecular-scale systems described by the overdamped Langevin equation, a first-order stochastic differential equation, recently developed
techniques make it possible to efficiently reconstruct the dynamics
from observed trajectories~\cite{Siegert1998a,Ragwitz2001,Beheiry2015,PerezGarcia2018,Frishman2018}.
Many complex systems at larger scales, however, exhibit stochastic dynamics governed by the \emph{underdamped} Langevin equation, a second-order stochastic differential equation. 
Examples include cell
motility~\cite{Selmeczi2005,Li2011,Sepulveda2013,Dalessandro2017,Brueckner2019}, 
postural dynamics in animals~\cite{Stephens2008,Stephens2011},
movement in interacting swarms of fish~\cite{Gautrais2012,Katz2011,Jhawar2020},
birds~\cite{Cavagna2010,Attanasi2014b}, and
insects~\cite{Buhl2006,Attanasi2014a}, as well as dust particles in a
plasma~\cite{Gogia2017}.
Due to recent advances in tracking
technology, the diversity, accuracy, dimensionality, and size of these
behavioral data-sets is rapidly
increasing~\cite{Brown2018}, resulting in a growing
need for accurate inference approaches for high-dimensional underdamped stochastic systems.
However, there is currently no rigorous method to infer the dynamics of such underdamped stochastic systems.

Inference from underdamped stochastic systems suffers from a major
challenge absent in the overdamped case. In any realistic application, the accelerations
of the degrees of freedom must be obtained as discrete second
derivatives from the observed position trajectories, which are sampled
at discrete intervals $\Delta t$. Consequently, a straightforward generalization of the estimators for the force and noise fields of overdamped systems fails: these estimators do not converge to the correct values, even in the limit $\Delta t \to 0$~\cite{Pedersen2016,Ferretti2019}. To make matters worse, real
data is always subject to measurement errors, leading to divergent
biases in the discrete estimators~\cite{Lehle2015}. These problems have so far precluded
reliable inference in underdamped stochastic systems.

Here, we introduce a general framework, Underdamped Langevin Inference (ULI), that conceptually explains the origin of these
 biases, and provides an operational scheme to reliably infer the equation of motion of underdamped stochastic systems governed by non-linear force fields and multiplicative noise amplitudes. To
provide a method that can be robustly applied to realistic
experimental data, we rigorously derive estimators that converge to
the correct values for discrete data subject to measurement errors. We demonstrate the power of our
method by applying it to experimental trajectories of single
migrating cells, as well as simulated complex high-dimensional data
sets, including flocks of active particles with Viscek-style alignment
interactions.

We consider a general $d$-dimensional stationary stochastic process $\mathbf{x}(t)$ with components $\{ x_\mu(t) \}_{1 \leqslant \mu \leqslant d}$ governed by the underdamped Langevin equation
\begin{align}
\label{eqn:process}
\begin{split}
\dot{x}_\mu &= v_\mu \\
\dot{v}_\mu &= F_\mu(\mathbf{x},\mathbf{v}) + \sigma_{\mu \nu}(\mathbf{x},\mathbf{v}) \xi_\nu(t)
\end{split}
\end{align}
which we interpret in the It\^o-sense. Throughout, we employ the Einstein summation convention, and $\xi_\mu(t)$ represents a Gaussian white noise with the properties $\langle \xi_\mu(t) \xi_\nu(t') \rangle = \delta_{\mu\nu}\delta(t-t')$ and $\langle \xi_\mu(t)\rangle =0$. Our aim is to infer the force field $F_\mu(\mathbf{x},\mathbf{v})$ and the noise amplitude $\sigma_{\mu \nu}(\mathbf{x},\mathbf{v})$ from an observed finite trajectory of the process~\footnote{Since we interpret eqn.~\eqref{eqn:process} in the It\^o-sense, the inferred force field $F_\mu(\mathbf{x},\mathbf{v})$ corresponds to this convention.}. 


\begin{figure*}[ht]
\centering
	\includegraphics[width=\textwidth]{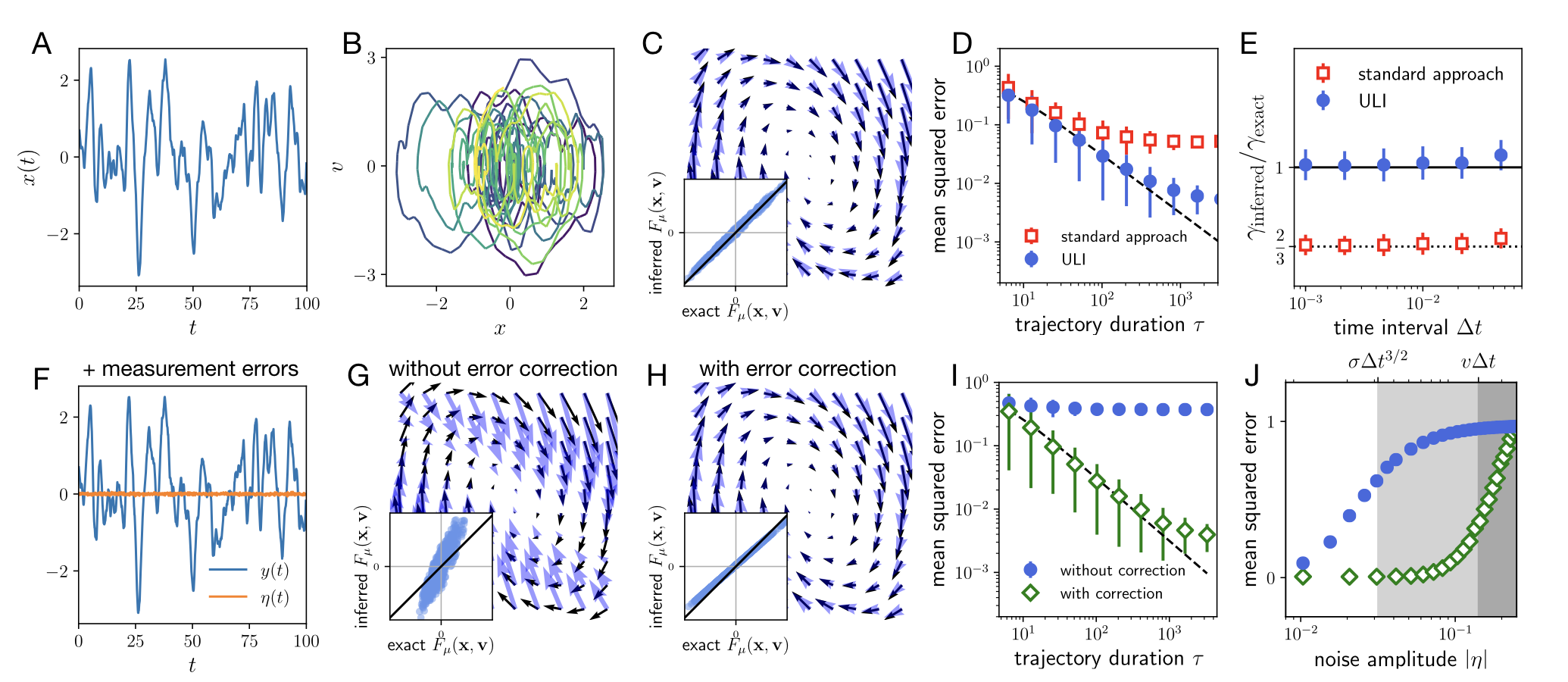}
	\caption{
		\textbf{Inference from discrete time series subject to measurement error.} 
		\textbf{A.} Trajectory $x(t)$ of a stochastic damped harmonic oscillator, $F(x,v)=-\gamma v - kx$.
		\textbf{B.} The same trajectory represented in $xv$-phase space. Color coding indicates time.
		\textbf{C.} Force field in $xv$-space inferred from the trajectory in A using ULI with basis functions $b = \{ 1, x, v \}$ (blue arrows), compared to the exact force  field (black arrows). \textit{Inset:} inferred components of the force along the trajectory \textit{versus} the exact values.
		\textbf{D.} Convergence of the mean squared error of the inferred force field, obtained using ULI (circles) and with the previous standard approach~\cite{Brueckner2019,Stephens2008,Pedersen2016,Lehle2015} (squares). Dashed lines indicate the predicted error $\delta \hat{F}^2/\hat{F}^2 \sim N_b/2\hat{I}_b$.
		\textbf{E.} Inferred friction coefficient $\gamma$ divided by the exact one as a function of the sampling time interval $\Delta t$, comparing the previous standard approach to ULI.
		\textbf{F.} Trajectory $y(t) = x(t) + \eta(t)$ (blue) corresponding to the same realization $x(t)$ in A, with additional time-uncorrelated measurement error $\eta(t)$ (orange) with small amplitude $|\eta|=0.02$. 		
		\textbf{G,H.} Force field inferred from $y(t)$ using estimators without and with measurement error corrections, respectively.
		\textbf{I.} Inference convergence for data subject to measurement error using estimators without (circles) and with (diamonds) measurement error corrections.
		\textbf{J.} Dependence of the inference error on the noise amplitude $|\eta|$ (same symbols as in I).
		}
	\label{fig1}
\end{figure*}


We start by approximating the force field as a linear combination of $n_b$ basis functions $b = \{  b_\alpha(\mathbf{x},\mathbf{v})  \}_{1 \leqslant \alpha \leqslant n_b}$, such as polynomials, Fourier modes, wavelet functions, or Gaussian kernels~\cite{Stephens2008}. From these basis functions, we construct an empirical orthonormal basis $\hat{c}_\alpha(\mathbf{x},\mathbf{v}) = \hat{B}_{\alpha \beta}^{-1/2}b_\beta(\mathbf{x},\mathbf{v})$ such that $\langle \hat{c}_\alpha \hat{c}_\beta \rangle = \delta_{\alpha \beta}$, an approach that was recently proposed for overdamped systems~\cite{Frishman2018}. Here and throughout, averages correspond to time-averages along the trajectory. We can then approximate the force field as $F_\mu(\mathbf{x},\mathbf{v}) \approx F_{\mu \alpha} \hat{c}_\alpha(\mathbf{x},\mathbf{v})$. Similarly, we perform a basis expansion of the noise amplitude $\sigma^2_{\mu \nu}(\mathbf{x},\mathbf{v})$. Thus, the inference problem reduces to estimating the projection coefficients $F_{\mu \alpha}$ and $\sigma^2_{\mu \nu \alpha}$.

\textbf{Dealing with discreteness --} In practice, only the
configurational coordinate $\mathbf{x}(t)$ is accessible in
experimental data, sampled at a discrete time-interval $\Delta t$. We
therefore only have access to the discrete estimators of the velocity
$\mathbf{\hat{v}}(t)= [\mathbf{x}(t) - \mathbf{x}(t-\Delta t)]/\Delta
t$ and acceleration
$\mathbf{\hat{a}}(t) = [\mathbf{x}(t+\Delta t) - 2\mathbf{x}(t) +
\mathbf{x}(t-\Delta t)]/{\Delta t^2}$. Our goal is to derive an
estimator $\hat{F}_{\mu \alpha}$, constructed from the discrete
velocities and accelerations, which converges to the exact projections
$F_{\mu \alpha}$ in the limit $\Delta t \rightarrow 0$.

An intuitive approach would be to simply generalize the estimators for
overdamped systems~\cite{Frishman2018} and calculate the projections
of the accelerations
$\langle \hat{a}_\mu \hat{c}_\alpha(\mathbf{x},\mathbf{\hat{v}})
\rangle$. This expression has indeed previously been used for
underdamped
systems~\cite{Brueckner2019,Stephens2008,Pedersen2016,Lehle2015}. We derive the correction term to
this estimator by expanding the basis functions
$\hat{c}_\alpha(\mathbf{x},\mathbf{\hat{v}}) =
\hat{c}_\alpha(\mathbf{x},\mathbf{v}) + (\partial_{v_\mu}
\hat{c}_\alpha) (\hat{v}_\mu - v_\mu) + ...$, where the leading order
contribution to the second term is a fluctuating (zero average) term
of order $\Delta t^{1/2}$. Similarly, we perform a stochastic
It\^o-Taylor expansion of the discrete acceleration
$\mathbf{\hat{a}}(t)$, which has a leading order fluctuating term of
order $\Delta t^{-1/2}$. Thus, while each of these terms individually
averages to zero, their product results in a bias term with non-zero
average of order $\Delta t^{0}$:
$\langle \hat{a}_\mu \hat{c}_\alpha(\mathbf{x},\mathbf{\hat{v}})
\rangle = F_{\mu\alpha} + \frac{1}{6} \left\langle \sigma^2_{\mu \nu}
  \partial_{v_\nu}c_\alpha(\mathbf{x},\mathbf{v}) \right\rangle +
\mathcal{O}(\Delta t)$~\footnote{See Supplemental Material at [URL
  will be inserted by publisher] for detailed derivations of the
  correction terms and estimators.}. As expected, this bias vanishes
in the limit $\sigma \to 0$, and therefore does not appear in
deterministic systems. However, it poses a problem wherever a second derivative of a stochastic
signal is averaged conditioned on its first derivative. The occurrence of such a bias was observed in linear
systems~\cite{Pedersen2016,Ferretti2019}. Specifically, for a linear
viscous force $F(v)=-\gamma v$, it was found that
$\langle \hat{a} c(\hat{v}) \rangle = -\frac{2}{3}\gamma +
\mathcal{O}(\Delta t)$, which is recovered by our general expression for the systematic bias~\cite{Note1}. 

Previous approaches to correct for this bias rely on \textit{a priori} knowledge of the observed stochastic
process~\cite{Pedersen2016}, are limited to simple parametric forms~\cite{Ferretti2019}, or perform an \textit{a
  posteriori} empirical iterative scheme~\cite{Brueckner2019}. In
contrast, by simply deducting the general form of the bias, we obtain
our Underdamped Langevin Inference (ULI) estimator~\cite{Note1}:
\begin{align}
\label{eqn:guesstimator}
\hat{F}_{\mu \alpha} 
	= \langle \hat{a}_\mu \hat{c}_\alpha(\mathbf{x},\mathbf{\hat{v}}) \rangle -
		 \frac{1}{6} \left\langle \widehat{\sigma^2}_{\mu \nu}(\mathbf{x},\mathbf{\hat{v}}) \partial_{v_\nu} \hat{c}_\alpha(\mathbf{x},\mathbf{\hat{v}}) \right\rangle
\end{align}
The presence of the derivative of a basis function in the estimator highlights the importance of projecting the dynamics of underdamped systems onto a set of \textit{smooth} basis functions, in contrast to the traditional approach of taking conditional averages in a discrete set of bins~\cite{Siegert1998a,Ragwitz2001}, equivalent to a basis of non-differentiable top-hat functions. 

Similarly to the force field, we expand the noise amplitude as a sum of basis functions, and derive an unbiased estimator for the projection coefficients~\cite{Note1}
\begin{align}
\label{eqn:difftimator}
\widehat{\sigma^2}_{\mu \nu \alpha} = \frac{3 \Delta t}{2} \langle \hat{a}_\mu \hat{a}_\nu \hat{c}_\alpha(\mathbf{x},\mathbf{\hat{v}}) \rangle
\end{align}
To test our method, we start with a simulated minimal example, the stochastic damped harmonic oscillator $\dot{v}=-\gamma v - kx+\sigma \xi$ (Fig.~\ref{fig1}A-E). Indeed, we find that even for such a simple system, the intuitive acceleration projections $\langle \hat{a}_\mu \hat{c}_\alpha(\mathbf{x},\mathbf{\hat{v}}) \rangle$ yield a biased result (Fig.~\ref{fig1}E). In contrast, ULI, defined by Eqs.~\eqref{eqn:difftimator} and~\eqref{eqn:guesstimator}, provides an accurate reconstruction of the force field (Fig.~\ref{fig1}C,E). To test the convergence of these estimators in a quantitative way, we calculate the expected random error due to the finite length $\tau$ of the input trajectory, 
$\delta \hat{F}^2/\hat{F}^2 \sim N_b/2\hat{I}_b$, 
where we define $\hat{I}_b = \frac{\tau}{2}\hat{\sigma}_{\mu\nu}^{-2}\hat{F}_{\mu\alpha}\hat{F}_{\nu\alpha}$ as the empirical estimate of the information contained in the trajectory, and $N_b=d n_b$ is the number of degrees of freedom in the force field~\cite{Frishman2018}. We confirm that the convergence of our estimators follows this expected trend, in contrast to the biased acceleration projections (Fig.~\ref{fig1}D). Therefore, ULI provides an operational method to accurately infer the dynamical terms of underdamped stochastic trajectories.

\textbf{Treatment of measurement errors $-$} A key challenge in stochastic inference from real data is the unavoidable presence of time-uncorrelated random measurement errors $\boldsymbol{\eta}(t)$, which can be non-Gaussian: the observed signal in this case is $\mathbf{y}(t) = \mathbf{x}(t) + \boldsymbol{\eta}(t)$. This problem is particularly dominant in underdamped inference, where the signal is differentiated twice, leading to a divergent bias of order $\Delta t^{-3}$~\cite{Note1}. Thus, for small $\Delta t$, even small measurement errors can lead to prohibitively large systematic inference errors, which cannot be rectified by simply recording more data.

To overcome this challenge, we derive estimators which are robust against measurement error. These estimators are constructed such that the leading-order bias terms cancel. For the force estimator, we find that this is achieved by using the local average position $\mathbf{\overline{x}}(t)=\frac{1}{3}(\mathbf{x}(t-\Delta t) + \mathbf{x}(t) +  \mathbf{x}(t+\Delta t))$ and the symmetric velocity $\mathbf{\hat{v}}(t)= [\mathbf{x}(t+\Delta t) - \mathbf{x}(t-\Delta t)]/(2\Delta t)$ in Eq.~\eqref{eqn:guesstimator}~\footnote{Note that due to the change of definition of $\mathbf{\hat{v}}$, the prefactor of the correction term in Eq.~\eqref{eqn:guesstimator} changes from $1/6$ to $1/2$.}. Similarly, we derive an unbiased estimator for the noise term, which is constructed using a linear combination of four-point increments~\cite{Note1}.

Remarkably, these modifications result in a vastly improved inference performance in the presence of measurement error (Fig.~\ref{fig1}F-J). Specifically, while the bias becomes dominant at an error magnitude $|\eta| \sim \sigma \Delta t^{3/2}$ in the standard estimators, the bias-corrected estimators only fail when the measurement error becomes comparable to the displacement in a single time-step, $|\eta| \sim v \Delta t$ (Fig.~\ref{fig1}J)~\cite{Note1}. Thus, our method has a significantly larger range of validity extending up to the typical displacement in a single time-frame. 

\begin{figure}[]
	\includegraphics[width=0.5\textwidth]{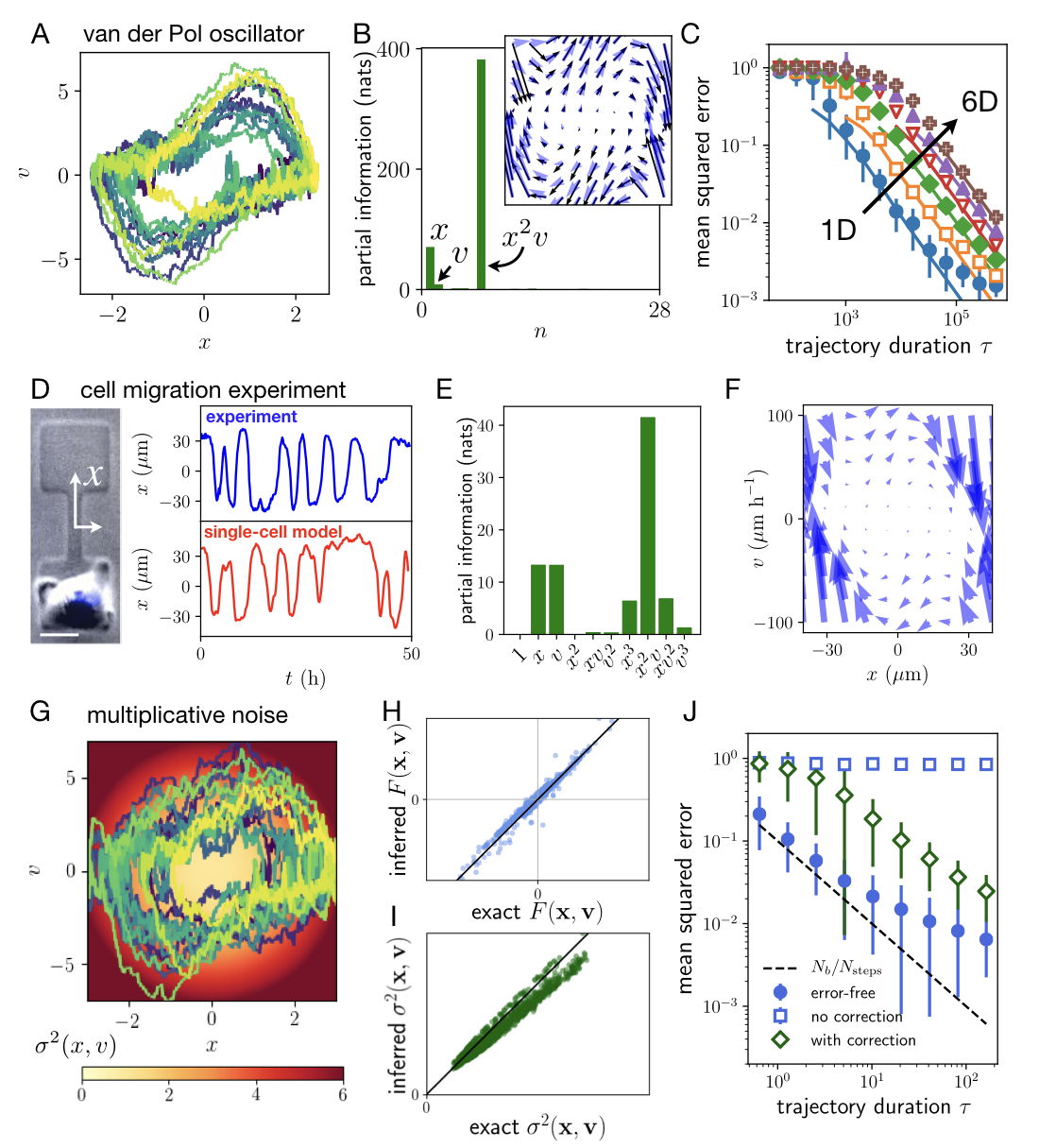}
	\caption{
		\textbf{Inferring non-linear dynamics and multiplicative noise.} 
		\textbf{A.} $xv$-trajectory of the stochastic Van der Pol oscillator, $F(x,v)=\kappa(1-x^2)v-x$ with measurement error.
		\textbf{B.} Partial information of the 28 basis functions of a 6th order polynomial basis in natural information units (1 nat $=1/\log2$ bits), inferred from the trajectory in A. \textit{Inset:} Corresponding force field reconstruction. 		
		\textbf{C.} Convergence of the inference error for the $d$-dimensional Van der Pol oscillator $F_\mu(\mathbf{x},\mathbf{{v}})=\kappa_\mu(1-x_\mu^2)v_\mu-x_\mu$ (no summation, $1 \leqslant \mu \leqslant d$) with $d=1...6$, using a third-order polynomial basis.
		\textbf{D.} Microscopy image of a migrating human breast cancer cell (MDA-MB-231) confined in a two-state micropattern (scale bar: $20 \mu$m). Experimental trajectory of the cell nucleus position, recorded at a time-interval $\Delta t = 10$ min (blue), and simulated trajectory using the inferred model (red). 
		\textbf{E.} Partial information for the experimental trajectory in D, projected onto a third-order polynomial basis.
		\textbf{F.} Deterministic flow field inferred from the experimental trajectory in D.		
		\textbf{G.} Trajectory of a Van der Pol oscillator with multiplicative noise $\sigma^2(x,v) = \sigma_0 + \sigma_x x^2+\sigma_v v^2$ (colormap).
		\textbf{H,I.} Inferred \textit{versus} exact components of the force and noise term, respectively, for the trajectory in G.
		\textbf{J.} Inference convergence of the multiplicative noise amplitude, using Eq.~\eqref{eqn:difftimator} without measurement error (circles), with measurement error (squares), and using the error-corrected estimator (diamonds). The error saturation at large $\tau$ is due to the finite time-step. Dashed line: predicted error $\delta \widehat{\sigma^2}/\widehat{\sigma^2} \sim \sqrt{N_b \Delta t/\tau}$~\cite{Frishman2018}.
		}
	\label{fig2}
\end{figure}

\textbf{Non-linear dynamics $-$} Since our method does not assume linearity, we can expand the projection basis to include higher order functions to capture the behavior of systems with non-linear dynamics. As a canonical example, we study the stochastic Van der Pol oscillator $\dot{v}=\kappa(1-x^2)v-x + \sigma \xi$, a common model for a broad range of biological dynamical systems~\cite{Kruse2005b}. We simulate a short trajectory of this process, with added artificial measurement error (Fig.~\ref{fig2}A). Indeed, we find that ULI reliably infers the underlying phase-space flow (Fig.~\ref{fig2}B). This is not limited to one-dimensional systems, as shown by studying convergence of higher-dimensional oscillators (Fig.~\ref{fig2}C). Importantly, this good performance does not rely on using a polynomial basis to fit a polynomial field: employing a non-adapted basis, such as Fourier components, yields similarly good results~\cite{Note1}. 

To capture the Van der Pol dynamics, only the three basis functions
$\{ x,v, x^2v\}$ are required. But can these functions be identified
directly from the data without prior knowledge of the underlying
force field? To address this question, we introduce the concept of partial
information. We can estimate the information contained in a finite
trajectory as
$\hat{I}_b(n_b) =
\frac{\tau}{2}\hat{\sigma}_{\mu\nu}^{-2}\hat{F}_{\mu\alpha}\hat{F}_{\nu\alpha}$,
where $\hat{F}_{\nu\alpha}$ are the projection coefficients onto the
basis $b$ with $n_b$ basis functions~\cite{Frishman2018}. To assess
the importance of the $n^\mathrm{th}$ basis function in the expansion,
we calculate the amount of information it contributes:
\begin{equation}
\label{eqn:part_info}
\hat{I}_b^{\mathrm{(partial)}}(n) = \hat{I}_b(n) - \hat{I}_b(n-1)
\end{equation}
which we term the partial information contributed by the basis function $b_n$. This approach successfully recovers the relevant terms in large basis sets (Inset Fig.~\ref{fig2}B). Thus, the partial information provides a useful heuristic for detecting the relevant terms of the force field.

To illustrate that ULI is practical and data-efficient, we apply it to experimental trajectories of cells migrating in two-state confinements (Fig.~\ref{fig2}D). Within their lifetime, these cells perform several transitions between the two states, resulting in relatively short trajectories. Previously, we inferred dynamical properties by averaging over a large ensemble of trajectories~\cite{Brueckner2019,Fink2019,Brueckner2020}. In contrast, with ULI, we can reliably infer the governing equation of motion from single cell trajectories. Here, $F(x,v)$ corresponds to the deterministic dynamics of the system, and not to a physical force. We employ the partial information to guide our basis selection: indeed, it recovers the intrinsic symmetry of the system, suggesting a symmetrized third order polynomial expansion is a suitable choice (Fig.~\ref{fig2}E). Using this expansion, we infer the deterministic flow field of the system (Fig.~\ref{fig2}F), which predicts trajectories similar to the experimental ones (Fig.~\ref{fig2}D). Importantly, the inferred model is self-consistent: re-inferring from short simulated trajectories yields a similar model~\cite{Note1}. Using ULI, we can thus perform inference on small data sets, enabling "single-cell profiling'', which could provide a useful tool to characterize cell-to-cell variability~\cite{Brueckner2020}.

To demonstrate the broad applicability of our approach, we evaluate its performance in the presence of multiplicative noise amplitudes $\sigma_{\mu \nu}(\mathbf{x},\mathbf{v})$, which occur in a range of complex systems~\cite{Friedrich2000,Brueckner2019,Stephens2008}. ULI accurately recovers the space- and velocity-dependence of both the force and noise field, and the estimators converge to the exact values, even in the presence of measurement errors (Fig.~\ref{fig2}G-J). To summarize, we have shown that ULI performs well on short trajectories of non-linear data sets subject to measurement errors, and can accurately infer the spatial structure of multiplicative noise terms. 

\begin{figure}
	\includegraphics[width=0.45\textwidth]{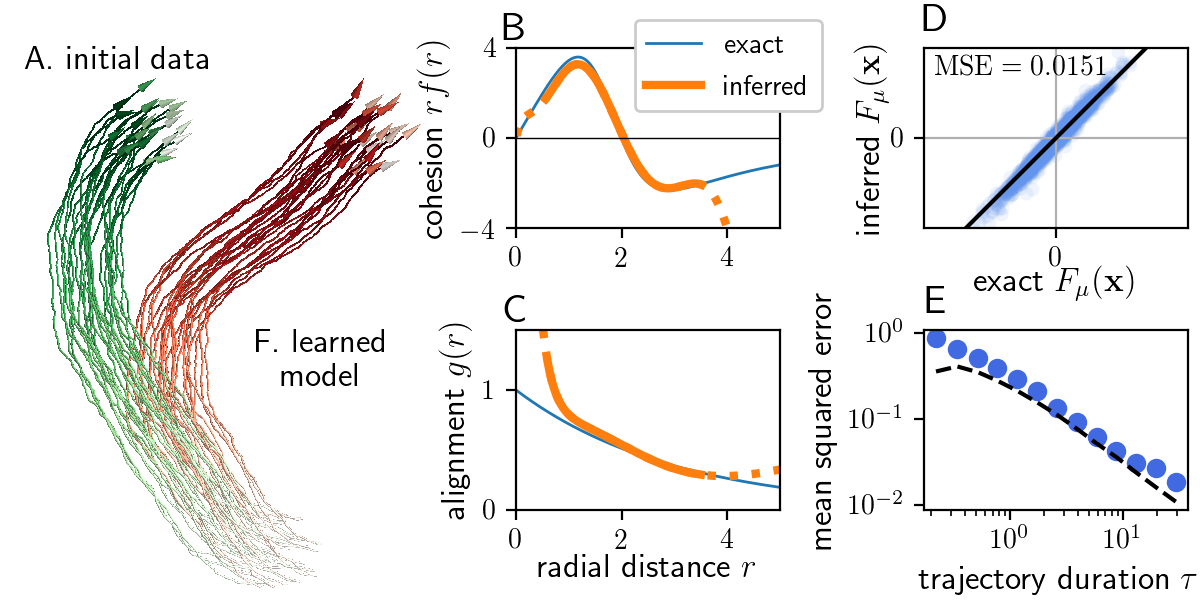}
	\caption{
		\textbf{Interacting flocks.} 
		\textbf{A.} Trajectory (green) of $N=27$ Viscek-like particles (Eq.~\ref{eqn:aligning}) in the flocking regime (1000 frames). We perform ULI on this trajectory using a translation-invariant basis of pair interaction and alignment terms, both fitted with $n=8$ exponential kernels.
                \textbf{B.}~Exact (blue) and inferred (orange) cohesion $rf(r)$. Exact form includes short-range repulsion and long-range attraction, $f(r) = \epsilon_0 (1 - (r/r_0)^3)/((r/r_0)^6+1)$. Dotted inference dependence indicates distances not sampled by the initial data.
                \textbf{C.} Exact and inferred alignment kernel $g(r)$. Exact form: $g(r) = \epsilon_1 \exp(-r/r_1)$.
                \textbf{D.} Inferred \emph{versus} exact components of the force field.
                \textbf{E.} Convergence of the inferred force as a function of trajectory length. Dashed line is the predicted error $\delta \hat{F}^2/\hat{F}^2 \sim N_b/2\hat{I}_b$.
                \textbf{F.} Simulated trajectory (red) employing the inferred force and noise, showing qualitatively similar flocking behavior.
		}
	\label{fig3}
\end{figure}

\textbf{Collective systems $-$} A major challenge in stochastic inference is the treatment of interacting many-body systems. In recent years, trajectory data on active collective systems, such as collective cell migration~\cite{Sepulveda2013,Dalessandro2017} and animal groups~\cite{Cavagna2010,Lukeman2010a,Buhl2006,Attanasi2014a,Attanasi2014b}, have become readily available. Previous approaches to such systems frequently focus on the study of correlations~\cite{Cavagna2010,Cavagna2017,Bialek2012} or collision statistics~\cite{Katz2011,Lukeman2010a,Dalessandro2017}, but no general method for inferring their underlying dynamics has been proposed. The collective behavior of these systems, ranging from disordered swarms \cite{Attanasi2014a} to ordered flocking \cite{Cavagna2010}, is determined by the interplay of active self-propulsion, cohesive and alignment interactions, and noise. Thus, disentangling these contributions could provide key insights into the physical laws governing active collective systems.

We  consider a simple model for the dynamics of a 3D flock with Viscek-style alignment interactions~\cite{Viscek1995,Gregoire2003,Chate2008,Sepulveda2013}, \begin{align}
\label{eqn:aligning}
\mathbf{\dot{v}}_{i} = \mathbf{p}_i + \sum_{j\neq i} \left[ f(r_{ij})\mathbf{r}_{ij} + g(r_{ij})\mathbf{v}_{ij} \right]  + \sigma \boldsymbol{\xi}_i
\end{align}
where $\mathbf{v}_{i}=\dot{\mathbf{r}}_{i}$, $\mathbf{r}_{ij} = \mathbf{r}_{j}-\mathbf{r}_{i}$, $\mathbf{v}_{ij} = \mathbf{v}_{j}-\mathbf{v}_{i}$, and $\mathbf{p}_i = \gamma(v_0^2-|\mathbf{v}_{i}|^2)\mathbf{v}_{i}$ is a self-propulsion force acting along the direction of motion of each particle $i$. Here, $f$ and $g$ denote the strength of the cohesive and alignment interactions, respectively, as a function of inter-particle distance $r_{ij}$. This model exhibits a diversity of behaviors, including flocking (Fig.~\ref{fig3}A). Intuitively, one might expect that ULI should fail dramatically in such a system: a 3D swarm of $N$ particles has $6N$ degrees of freedom, and ``curse of dimensionality'' arguments make this problem seem intractable. However, by exploiting the particle exchange symmetry and radial symmetry of the interactions~\cite{Note1}, we find that ULI accurately recovers the cohesion and alignment terms (Fig.~\ref{fig3}B-C), and captures the full force field (Fig.~\ref{fig3}D,E). Furthermore, simulating the inferred model yields trajectories with high similarity to the input data (Fig.~\ref{fig3}F). This example illustrates the potential of ULI for inferring complex interactions from trajectories of stochastic many-body systems.

In summary, we demonstrate how to reliably infer the force and noise fields in complex underdamped stochastic systems. We show that the inevitable presence of discreteness and measurement errors result in systematic biases that have so far prohibited accurate inference. To circumvent these problems, we have rigorously derived unbiased estimators, providing an operational framework, Underdamped Langevin Inference, to infer underdamped stochastic dynamics~\footnote{A readily usable \textsc{Python} package to perform Underdamped Langevin Inference is available at https://github.com/ronceray/UnderdampedLangevinInference.}. Our method provides a new avenue to analyzing the dynamics of complex high-dimensional systems, such as assemblies of motile cells~\cite{Sepulveda2013,Dalessandro2017}, active swarms~\cite{Cavagna2010,Lukeman2010a,Buhl2006,Attanasi2014a}, as well as non-equilibrium condensed matter systems~\cite{Baldovin2019,Gogia2017,Kruse2005b}. \\

\begin{acknowledgments}
We thank Alexandra Fink and Joachim R\"adler for generously providing experimental cell trajectories. Funded by the Deutsche Forschungsgemeinschaft (DFG, German Research Foundation) - Project-ID 201269156 - SFB 1032 (Project B12). D.B.B. is supported by a DFG fellowship within the Graduate School of Quantitative Biosciences Munich (QBM) and by the Joachim Herz Stiftung. 
\end{acknowledgments}

\bibliography{../library}



\section*{Supplementary material}

# Supplementary Material:
## Inferring the dynamics of underdamped stochastic systems

David B. Brückner*, Pierre Ronceray* and Chase P. Broedersz

This Supplemental Material contains a detailed definition of the projection formalism we employ in Underdamped Langevin Inference (ULI) (section 1), derivations of the unbiased estimators for the force and noise fields for discrete data (section 2) and for discrete data with random measurement errors (section 3), a criterion to choose the optimal basis size $n_b$ (section 4), further details on the inference from experimental single cell trajectories (section 5) and detailed information on the models and parameters used for the simulation results shown in Figs. 1-3 in the main text (section 6).

## Contents

## 1 Definition of the projection formalism

We consider $d$-dimensional processes $\mathbf{x}(t)$, governed by the underdamped Langevin equation

$$\dot{x}_\mu = v_\mu$$
$$\dot{v}_\mu = F_\mu(\mathbf{x}, \mathbf{v}) + \sigma_{\mu\nu}(\mathbf{x}, \mathbf{v})\xi_\nu(t) \tag{S1}$$

Here, $x_\mu(t)$ are the components of the vector $\mathbf{x}(t)$ and $1 \leqslant \mu \leqslant d$. The term $F_\mu(\mathbf{x}, \mathbf{v})$ denotes the $\mu$-component of the force field, and $\sigma_{\mu\nu}(\mathbf{x}, \mathbf{v})$ is the noise strength tensor, which is multiplicative

and can thus depend on the state of the system given by $\{\mathbf{x}, \mathbf{v}\}$. $\xi_\mu(t)$ represents a Gaussian white noise with the properties $\langle \xi_\mu(t)\xi_\nu(t') \rangle = \delta_{\mu\nu}\delta(t-t')$ and $\langle \xi_\mu(t) \rangle = 0$. Our aim is to infer the force field $F_\mu(\mathbf{x}, \mathbf{v})$ and the noise amplitude $\sigma_{\mu\nu}(\mathbf{x}, \mathbf{v})$ from an observed trajectory of the process. We interpret this stochastic differential equation in the Itô-sense, and thus infer the force field $F_\mu(\mathbf{x}, \mathbf{v})$ corresponding to this convention, which may include spurious drift terms due to multiplicative noise amplitudes.

In ULI, we approximate the force field as a linear combination of basis functions $b = \{b_\alpha(\mathbf{x}, \mathbf{v})\}$ where the index $\alpha$ runs over all basis functions in the set, $1 \leqslant \alpha \leqslant n_b$. Here and throughout the main text and supplementary material, we employ the Einstein summation convention. Thus, summations over the basis functions, indexed by $\{\alpha, \beta\}$ run from $1...n_b$. Summations over the $d$-dimensional dynamical quantities such as $x_\mu(t), v_\mu(t), F_\mu(\mathbf{x}, \mathbf{v}), \sigma_{\mu\nu}(\mathbf{x}, \mathbf{v})$, indexed by $\{\mu, \nu, \rho, \tau, ...\}$, run from $1...d$.

To extract the coefficients of the expansions of the force and noise terms, we can project the dynamics onto the space spanned by $b_\alpha(\mathbf{x}, \mathbf{v})$ using the steady-state probability distribution $P(\mathbf{x}, \mathbf{v})$ as a measure [1]. To do so, we define orthonormalized projectors $c_\alpha(\mathbf{x}, \mathbf{v}) = B_{\alpha\beta}^{-1/2} b_\beta(\mathbf{x}, \mathbf{v})$, such that

$$\langle c_\alpha c_\beta \rangle = \int c_\alpha(\mathbf{x}, \mathbf{v}) c_\beta(\mathbf{x}, \mathbf{v}) P(\mathbf{x}, \mathbf{v}) \mathrm{d}\mathbf{x}\mathrm{d}\mathbf{v} = \delta_{\alpha\beta} \quad (S2)$$

We then approximate the force field as a linear combination of these basis functions

$$F_\mu(\mathbf{x}, \mathbf{v}) \approx F_{\mu\alpha} c_\alpha(\mathbf{x}, \mathbf{v}) \quad (S3)$$

Note that if we use a complete set of basis functions, this becomes an exact equality. In any real application however, a truncated set of basis functions must be used, in order to limit the number of parameters (*i.e.* the coefficients $F_{\mu\alpha}$) to be inferred from a trajectory of finite length. The projection coefficients $F_{\mu\alpha}$ are given by

$$F_{\mu\alpha} = \int F_\mu(\mathbf{x}, \mathbf{v}) c_\alpha(\mathbf{x}, \mathbf{v}) P(\mathbf{x}, \mathbf{v}) \mathrm{d}\mathbf{x}\, \mathrm{d}\mathbf{v} \quad (S4)$$

These coefficients thus form a $(d \times n_b)$ matrix, and $F_{\mu\alpha}$ gives the projection coefficient of the $\mu$-component of the force field onto the basis function $c_\alpha$. Similarly, we expand the noise term

$$\sigma_{\mu\nu}^2(\mathbf{x}, \mathbf{v}) \approx \sigma_{\mu\nu\alpha}^2 c_\alpha(\mathbf{x}, \mathbf{v}) \quad (S5)$$

with the projection coefficients

$$\sigma_{\mu\nu\alpha}^2 = \int \sigma_{\mu\nu}^2(\mathbf{x}, \mathbf{v}) c_\alpha(\mathbf{x}, \mathbf{v}) P(\mathbf{x}, \mathbf{v}) \mathrm{d}\mathbf{x}\, \mathrm{d}\mathbf{v} \quad (S6)$$

Note that we expand $\sigma^2$ rather than $\sigma$ because we can only derive estimators for $\sigma^2$; since the noise averages to zero, we must take squares of the increments to extract the magnitude of the fluctuations.

In practice, we aim to infer the force and noise fields governing the dynamics of a system from a single trajectory of finite length $\tau$, sampled at a time interval $\Delta t$. Thus, the exact probability distribution $P(\mathbf{x}, \mathbf{v})$ is unknown, and we cannot enforce the condition Eq. (S2) exactly. Thus, we define *empirical* orthonormalized projectors

$$\hat{c}_\alpha(\mathbf{x}, \mathbf{v}) = \hat{B}_{\alpha\beta}^{-1/2} b_\beta(\mathbf{x}, \mathbf{v}) \quad (S7)$$

where
$$\hat{B}_{\alpha\beta} = \frac{\Delta t}{\tau} \sum_t b_\alpha(\mathbf{x}(t), \mathbf{v}(t)) b_\beta(\mathbf{x}(t), \mathbf{v}(t)), \tag{S8}$$

such that $\langle \hat{c}_\alpha \hat{c}_\beta \rangle = \delta_{\alpha\beta}$ where $\langle ... \rangle$ refers to a time-average along the observed trajectory.

Our aim in performing inference is to find the terms $F_\mu(\mathbf{x}, \mathbf{v})$ and $\sigma^2_{\mu\nu}(\mathbf{x}, \mathbf{v})$. Thus, we search for an operational definition of the estimators of the projection coefficients, $\hat{F}_{\mu\alpha}$ and $\widehat{\sigma^2}_{\mu\nu\alpha}$. These estimators consists of increment-constructions projected onto the trajectory-dependent orthonormal basis functions, constructed in such a way that the leading order term in $\Delta t$ converge to the exact projection coefficients. Due to the Gaussian nature of the stochastic noise, this projection procedure – which is equivalent to a least-square regression of the local estimator – corresponds for the force field to a maximum-likelihood approximation [1].

## 2 Derivation of the discrete estimators

To derive the leading order bias in the estimators for $F$ and $\sigma$, we start by defining the increments of the positions:
$$\Delta x_\mu^{(n)}(t) = x_\mu(t + n\Delta t) - x_\mu(t) \tag{S9}$$

The estimator for the accelerations is then given by a linear combination of these increments:
$$\hat{a}_\mu(t) = \frac{\Delta x_\mu^{(2)}(t) - 2\Delta x_\mu^{(1)}(t)}{\Delta t^2} = \frac{x_\mu(t + 2\Delta t) - 2x_\mu(t + \Delta t) + x_\mu(t)}{\Delta t^2} \tag{S10}$$

Note that this is not in general the most natural way to define $\hat{a}_\mu(t)$, as this expression is not centered around $t$. However, it makes the expression causal: the noise at $t' > t$ is independent of the state at $t$, thus using forward increments significantly simplifies the calculations. We will later shift the definition back to a centered one, which will only add higher order terms to our results. Similarly, we define the discrete velocity estimator

$$\hat{v}_\mu(t; \lambda) = \frac{\lambda \Delta x_\mu^{(2)}(t)}{2\Delta t} + \frac{(1 - \lambda) \Delta x_\mu^{(1)}(t)}{\Delta t} \tag{S11}$$

Note that we have kept some freedom in how we calculate the velocities from the three points $\{t, t + \Delta t, t + 2\Delta t\}$, denoted by the parameter $\lambda$. While most previous approaches [2, 3, 4, 5] use $\lambda = 0$, we will later show that in the presence of measurement errors, we have to choose $\lambda = 1$ (i.e. $\hat{v}$ odd under time reversal around $t + \Delta t$) to obtain an unbiased estimator. For now, we keep it as a variable parameter.

## 2.1 Itô integrals

Throughout this appendix, we will make use of Itô integrals [6], defined as follows:

$$I_0^{(n)} = \int_t^{t+n\Delta t} ds = n\Delta t \tag{S12}$$

$$I_{00}^{(n)} = \int_t^{t+n\Delta t} ds \int_t^s ds' = (n\Delta t)^2 \tag{S13}$$

$$I_\mu^{(n)} = \int_t^{t+n\Delta t} d\xi_\mu(s) \tag{S14}$$

$$I_{0\mu}^{(n)} = \int_t^{t+n\Delta t} ds \int_t^s d\xi_\nu(s') \tag{S15}$$

$$I_{\mu\nu}^{(n)} = \int_t^{t+n\Delta t} d\xi_\mu(s) \int_t^s d\xi_\nu(s') \quad \text{etc.} \tag{S16}$$

Throughout the text, we will frequently make use of the following identity

$$\langle I_{0\mu}^{(n)} I_{0\nu}^{(m)} \rangle = (\Delta t^3)\delta_{\mu\nu} f_{nm} \quad \text{where} \quad f_{nm} = \begin{cases} 1/3 & n = m = 1 \\ 5/6 & n = 1, m = 2 \\ 8/3 & n = m = 2 \\ 4/3 & n = 1, m = 3 \\ 14/3 & n = 2, m = 3 \\ 9 & n = m = 3 \end{cases} \tag{S17}$$

## 2.2 Force field

A first intuitive guess for the estimator of the force projections $F_{\mu\alpha}$ are the average projections of the acceleration

$$\hat{A}_{\mu\alpha} = \langle \hat{a}_\mu c_\alpha(\mathbf{x}, \hat{\mathbf{v}}) \rangle \tag{S18}$$

and indeed, this quantity has been used as a proxy for $F_{\mu\alpha}$ throughout the literature [2, 3, 4, 5]. To rigorously derive the leading order contributions to this quantity in terms of the dynamical terms $F_\mu$ and $\sigma_{\mu\nu}$, we start by expanding the increments

$$\Delta x_\mu^{(n)} = v_\mu I_0^{(n)} + \int_t^{t+n\Delta t} ds (v_\mu(s) - v_\mu) \tag{S19}$$

$$= v_\mu I_0^{(n)} + \int_t^{t+n\Delta t} ds \left[ \int_t^s ds' F_\mu(\mathbf{x}(s'), \mathbf{v}(s')) + \int_t^s d\xi_\nu(s') \sigma_{\mu\nu}(\mathbf{x}(s'), \mathbf{v}(s')) \right] \tag{S20}$$

$$= v_\mu I_0^{(n)} + \sigma_{\mu\nu} I_{0\nu}^{(n)} + F_\mu I_{00}^{(n)} + (\partial_{v_\rho} \sigma_{\mu\nu}) \sigma_{\rho\tau} I_{0\nu\tau}^{(n)} + \mathcal{O}(\Delta t^{5/2}) \tag{S21}$$

where we defined $v_\mu \equiv v_\mu(t)$, $F_\mu \equiv F_\mu(\mathbf{x}(t), \mathbf{v}(t))$, etc. We will use this short-hand notation as well as the Einstein summation convention throughout.

Next, we expand the basis functions $c_\alpha(\mathbf{x}, \hat{\mathbf{v}})$ around the true velocities $\mathbf{v}$:

$$c_\alpha(\mathbf{x}, \hat{\mathbf{v}}) = c_\alpha(\mathbf{x}, \mathbf{v}) + (\partial_{v_\rho} c_\alpha)(\hat{v}_\rho - v_\rho) + \frac{1}{2}(\partial^2_{v_\rho v_\tau} c_\alpha)(\hat{v}_\rho - v_\rho)(\hat{v}_\tau - v_\tau) + \mathcal{O}(\Delta t^{3/2}) \tag{S22}$$

From Eq. (S21), the leading order term of $\hat{v}_\rho - v_\rho$ is given by

$$\hat{v}_\rho - v_\rho = \frac{\lambda}{2\Delta t}\sigma_{\rho\nu}I^{(1)}_{0\nu} + \frac{1-\lambda}{\Delta t}\sigma_{\rho\nu}I^{(2)}_{0\nu} + \mathcal{O}(\Delta t) \quad (S23)$$

Thus, the leading order contribution to the second term of Eq. (S23) is a fluctuating (zero average) term of order $\Delta t^{1/2}$. To evaluate Eq. (S18), we also need the acceleration estimator $\hat{a}_\mu$. Substituting Eq. (S21) into Eq. (S10), we find the leading order terms of the acceleration estimator

$$\hat{a}_\mu = \frac{1}{\Delta t^2}\left[\sigma_{\mu\nu}(I^{(2)}_{0\nu} - 2I^{(1)}_{0\nu}) + F_\mu \Delta t^2\right] + \mathcal{O}(\Delta t^{3/2}) \quad (S24)$$

Thus, the leading order contribution to the acceleration is a fluctuating (zero average) term of order $\Delta t^{-1/2}$.

When we evaluate Eq. (S18) by substituting Eq. (S21) and (S23), we obtain

$$\hat{A}_{\mu\alpha} = \langle F_\mu c_\alpha(\mathbf{x}, \mathbf{v})\rangle + \frac{1+2\lambda}{6}\langle (\partial_{v_\rho} c_\alpha(x, v))\sigma_{\rho\nu}\sigma_{\mu\nu}\rangle + \mathcal{O}(\Delta t) \quad (S25)$$

The second term in this expression is an $\mathcal{O}(\Delta t^0)$-bias which means that the acceleration projections do not converge to the projections of the force, even in the limit of infinite sampling rate ($\Delta t \to 0$). This cross-term originates from the product of the fluctuating terms in the basis functions (of order $\Delta t^{1/2}$) and the accelerations (of order $\Delta t^{-1/2}$), which multiplied together give a term of order $\Delta t^0$ with non-zero average.

Our expression for the $\mathcal{O}(\Delta t^0)$-bias has several interesting properties:

- As one might expect, it vanishes in the deterministic limit $\sigma \to 0$; it is thus a property of stochastic systems.

- It vanishes for purely positional terms in the force-field, as it depends on the derivative $\partial_{v_\rho} c_\alpha(x, v)$. This makes sense, since it originates from the $\hat{v}$-dependence of the basis functions (Eq. (S23)). As shown by our derivation, it is a consequence of averaging the second derivative of a stochastic signal conditioned on its first derivative.

- A seemingly simple solution to remove the bias would be to set $\lambda = -1/2$. This results in a rather unconventional definition of the discrete velocity estimator,

$$\hat{v}_\mu(t; \lambda = -1/2) = \frac{1}{\Delta t}\left[-\frac{1}{4}x(t+2\Delta t) + \frac{3}{2}x(t+\Delta t) - \frac{5}{4}x(t)\right] \quad (S26)$$

for which $\hat{A}_{\mu\alpha}$ is a convergent estimator of $F_{\mu\alpha}$. However, using this definition of $\hat{v}_\mu$ results in large correction terms at the next order in $\Delta t$, and thus does not perform well at finite $\Delta t$. This estimator would also be strongly biased by measurement errors. For these reasons, we disregard it and turn to the derivation of a better estimator.

For $\lambda \neq -1/2$, the bias does not vanish, and has to be explicitly corrected for. Eq. (S25) allows us to derive an estimator for $F_{\mu\alpha}$ which is unbiased to first order in $\Delta t$, i.e. which converges to the exact projection coefficients in the limit $\Delta t \to 0$:

$$\hat{F}_{\mu\alpha} = \langle \hat{a}_\mu c_\alpha(\mathbf{x}, \hat{\mathbf{v}})\rangle - \frac{1+2\lambda}{6}\left\langle (\partial_{v_\nu} c_\alpha(\mathbf{x}, \hat{\mathbf{v}}))\widehat{\sigma^2_{\mu\nu}}(\mathbf{x}, \hat{\mathbf{v}})\right\rangle \quad (S27)$$

Note that in going from Eq. (S25) to Eq. (S27), we have replaced $v$ and $\sigma$ by their estimators, as their values are not known. This introduces additional correction terms, but these are of higher order in $\Delta t$. Eq. (S27) further implies that the noise term $\sigma^2$ has to be inferred before the force field can be inferred. In the presence of measurement errors (section 3), we show below that we must choose $\lambda = 1$, rendering $\hat{v}_\mu$ odd under time reversal around $t + \Delta t$. We therefore use this choice for $\lambda$ throughout.

Note that Eq. (S27) now conditions the acceleration $\hat{a}_\mu$ (Eq. (S10)) on its first point $x(t)$. In order to make this estimator symmetric, we shift the conditioning $c(\mathbf{x}(t), \hat{\mathbf{v}}(t)) \to c(\mathbf{x}(t + \Delta t), \hat{\mathbf{v}}(t))$. The resulting corrections, due to expanding $c(\mathbf{x}(t + \Delta t), \hat{\mathbf{v}}(t))$ around $\mathbf{x}(t)$, are of higher order in $\Delta t$. We can then relabel all time points such that $t \to t - \Delta t$, to arrive at our final formula for the estimator:

$$\hat{F}_{\mu\alpha} = \langle \hat{a}_\mu c_\alpha(\mathbf{x}, \hat{\mathbf{v}}) \rangle - \frac{1}{2} \left\langle (\partial_{v_\nu} c_\alpha(\mathbf{x}, \hat{\mathbf{v}})) \widehat{\sigma^2}_{\mu\nu}(\mathbf{x}, \hat{\mathbf{v}}) \right\rangle \tag{S28a}$$

$$\mathbf{x} = \mathbf{x}(t) \tag{S28b}$$

$$\hat{\mathbf{v}} = \frac{\mathbf{x}(t + \Delta t) - \mathbf{x}(t - \Delta t)}{2\Delta t} \tag{S28c}$$

$$\hat{\mathbf{a}} = \frac{\mathbf{x}(t + \Delta t) - 2\mathbf{x}(t) + \mathbf{x}(t - \Delta t)}{\Delta t^2} \tag{S28d}$$

## 2.3 Noise term

To derive an estimator for $\sigma$, we derive the leading order contributions to the quantity

$$\Delta t \langle \hat{a}_\mu \hat{a}_\nu \hat{c}_\alpha(\mathbf{x}, \hat{\mathbf{v}}) \rangle = \Delta t \langle [\sigma_{\mu\rho} I^{(2)}_{0\rho} - 2\sigma_{\mu\rho} I^{(1)}_{0\rho}][\sigma_{\nu\rho} I^{(2)}_{0\rho} - 2\sigma_{\nu\rho} I^{(1)}_{0\rho}] c_\alpha(\mathbf{x}, \mathbf{v}) \rangle + \mathcal{O}(\Delta t) \tag{S29}$$

$$= \frac{2}{3} \langle \sigma_{\mu\rho} \sigma_{\nu\rho} c_\alpha(\mathbf{x}, \mathbf{v}) \rangle + \mathcal{O}(\Delta t) \tag{S30}$$

where we have used Eqs. (S17), (S21). Here, the somewhat counter-intuitive factor of 2/3 stems from the expectation values of the Itô-integrals given in Eq. (S17). Thus, an unbiased estimator to first order in $\Delta t$ for the noise term is

$$\widehat{\sigma^2}_{\mu\nu} = \frac{3\Delta t}{2} \langle \hat{a}_\mu \hat{a}_\nu \hat{c}_\alpha(\mathbf{x}, \hat{\mathbf{v}}) \rangle \tag{S31}$$

## 2.4 Comparison to the exact formula for a linear damping force

Pedersen et al. [3] calculated the discretization effect for a linear viscous damping force $F(v) = -\gamma v$ (*i.e.* the one-dimensional underdamped Ornstein-Uhlenbeck process; in its discrete form also know as the Persistent Random Walk), to which we can compare our expression for the $\mathcal{O}(\Delta t^0)$-bias. The equation of motion for this process is given by

$$\dot{v} = -\gamma v + \sigma \xi(t) \tag{S32}$$

In ref. [3], the acceleration projections

$$\langle \hat{a} | \hat{v} \rangle = -\hat{\gamma} \hat{v} \tag{S33}$$

are considered. Here, $\langle \hat{a}|\hat{v}\rangle$ denotes conditional averaging of $\hat{a}$ with respect to $\hat{v}$, which is equivalent to using a basis of $\delta$-functions, *i.e.* $b_\alpha(v) = \delta(v - v^{(\alpha)})$. Using our definition of the velocity estimator (Eq. (S11)) with $\lambda = 0$, one obtains [3]

$$\hat{\gamma} = \frac{1}{\Delta t}\left[1 - \frac{(1-e^{-\gamma\Delta t})^2}{2(e^{-\gamma\Delta t} - 1 + \gamma\Delta t)}\right] \approx \frac{2}{3}\gamma - \frac{5}{18}\gamma^2\Delta t + \frac{23}{270}\gamma^3\Delta t^2 + \mathcal{O}(\Delta t^3) \quad \text{(S34)}$$

From Eq. (S25), we expect to find a similar bias, since we are considering a $v$-dependent component of the force field. To compare Eq. (S34) to our result, we use the basis $b = \{v\}$. Then, the normalised projection coefficient is given by

$$c(v) = \frac{v}{\sqrt{\langle v^2 \rangle}} = \frac{\sqrt{2\gamma}}{\sigma}v \quad \text{(S35)}$$

since $\langle v^2 \rangle = \sigma^2/2\gamma$ for the Ornstein-Uhlenbeck process. Thus, Eq. (S25) predicts

$$\langle \hat{a}_\mu c_\alpha(\hat{v})\rangle = F_{\mu\alpha} + \frac{(\partial_v c)\sigma^2}{6} + \mathcal{O}(\Delta t) = F_{\mu\alpha} + \frac{\sqrt{2\gamma}\sigma}{6} + \mathcal{O}(\Delta t) \quad \text{(S36)}$$

and therefore

$$\langle \hat{a}_\mu c_\alpha(\hat{v})\rangle c_\alpha(v) = F_\mu + \frac{\gamma}{3}v + \mathcal{O}(\Delta t) = -\frac{2}{3}\gamma v + \mathcal{O}(\Delta t) \quad \text{(S37)}$$

Thus, our approach recovers the leading order correction of the expression derived by Pedersen et al. [3].

## 3 Derivation of estimators in the presence measurement errors

In any real experiment, the recorded positions are subject to measurement errors, due to, e.g., motion blur or uncorrelated localization errors. Such random measurement errors can be modelled as an uncorrelated noise $\eta_\mu(t)$ (not necessarily Gaussian) acting on the positions $x_\mu(t)$ [3, 7], meaning that the signal we actually observe is

$$y_\mu(t) = x_\mu(t) + \eta_\mu(t) \quad \text{(S38)}$$

where $\langle \eta_\mu(t)\eta_\nu(t')\rangle = \Lambda_{\mu\nu}\delta(t-t')$.

### 3.1 Force field

We will now again calculate the leading order contributions to the estimator of the projected accelerations with measurement error (w.m.e.):

$$\hat{A}_{\mu\alpha}^{(\text{w.m.e.})} = \langle \hat{a}_\mu^{(\text{w.m.e.})} c_\alpha(\bar{\mathbf{y}}, \hat{\mathbf{w}})\rangle \quad \text{(S39)}$$

Here, $\hat{\mathbf{a}}^{(\text{w.m.e.})}$ and $\hat{\mathbf{w}}$ are the empirical acceleration and velocity derived from the signal subject to measurement error $\mathbf{y}(t)$, respectively. Note that we are no longer conditioning on a single position-like coordinate, but rather the average quantity $\bar{\mathbf{y}}$, which is a linear combination of the

three time-points entering the acceleration. This allows us to find a conditioning in terms of $\bar{\mathbf{y}}$ and $\hat{\mathbf{w}}$ such that the leading order terms due to the measurement errors cancel. We thus write

$$\bar{y}_\mu(\beta, \gamma) = \beta y_\mu(t + 2\Delta t) + \gamma y_\mu(t) + (1 - (\beta + \gamma))y_\mu(t + \Delta t) \tag{S40}$$

The velocity estimator including measurement noise is

$$\hat{w}_\mu = \hat{v}_\mu + \frac{\lambda \Delta \eta_\mu^{(2)}}{2\Delta t} + \frac{(1-\lambda)\Delta \eta_\mu^{(1)}}{\Delta t} := \hat{v}_\mu + \frac{f_\mu^{(v)}(\eta; \lambda)}{\Delta t} \tag{S41}$$

Similarly,

$$\hat{a}_\mu^{(\text{w.m.e.})} = \hat{a}_\mu + \frac{\Delta \eta_\mu^{(2)} - \Delta \eta_\mu^{(1)}}{\Delta t^2} := \hat{a}_\mu + \frac{f_\mu^{(a)}(\eta)}{\Delta t^2} \tag{S42}$$

We assume here that the measurement error $\eta_\mu$ is relatively small compared to the scale of variation of the fitting functions, such that we can expand the basis functions as

$$c_\alpha(\bar{\mathbf{y}}, \hat{\mathbf{w}}) = c_\alpha(\bar{\mathbf{x}}, \hat{\mathbf{v}}) + (\partial_{x_\nu} c_\alpha)\bar{\eta}_\nu(\beta, \gamma) + (\partial_{v_\nu} c_\alpha)\frac{f_\nu^{(v)}(\eta; \lambda)}{\Delta t} + \mathcal{O}(\eta^2) \tag{S43}$$

Combining Eqs. (S43) and (S42), the estimator of the acceleration projection thus reads

$$\hat{A}_{\mu\alpha}^{(\text{w.m.e.})} = \hat{A}_{\mu\alpha} + (\partial_{x_\nu} c_\alpha)\frac{\langle \bar{\eta}_\nu(\beta, \gamma) f_\mu^{(a)}(\eta)\rangle}{\Delta t^2} + (\partial_{v_\nu} c_\alpha)\frac{\langle f_\nu^{(v)}(\eta; \lambda) f_\mu^{(a)}(\eta)\rangle}{\Delta t^3} + \mathcal{O}(\eta^2) \tag{S44}$$

This shows that the leading order contribution to the estimator of the acceleration projection is of order $\Delta t^{-3}$, inducing a "dangerous" bias which diverges fast with $\Delta t \to 0$.

Indeed, the standard approach [2, 3, 4, 5] is to take $\hat{A}_{\mu\alpha}^{(\text{w.m.e.})}$ with $\lambda = \beta = \gamma = 0$ as a proxy for the force projections, which results in

$$\hat{A}_{\mu\alpha}^{(\text{w.m.e.})} = \hat{A}_{\mu\alpha} - (\partial_{x_\nu} c_\alpha)\frac{2\Lambda_{\mu\nu}}{\Delta t^2} - (\partial_{v_\nu} c_\alpha)\frac{3\Lambda_{\mu\nu}}{\Delta t^3} + \mathcal{O}(\eta^2) \tag{S45}$$

Here we propose to make use of the free parameters $\lambda$, $\beta$ and $\gamma$, to find a construction for our estimator such that the divergent cross-terms in Eq. (S44) cancel. Thus, we solve the following equations for $\{\lambda, \beta, \gamma\}$:

$$\langle \bar{\eta}_\nu(\beta, \gamma) f_\mu^{(a)}(\eta)\rangle = [(\beta + \gamma) - 2(1 - (\beta + \gamma))]\Lambda_{\mu\nu} \tag{S46}$$

$$\langle f_\nu^{(v)}(\eta; \lambda) f_\mu^{(a)}(\eta)\rangle = [3\lambda - 3]\Lambda_{\mu\nu} \tag{S47}$$

These terms vanish for $\lambda = 1$ and $\beta + \gamma = 2/3$. There is thus a remaining freedom in the choice of $\beta$ and $\gamma$. For simplicity, we choose the symmetric option $\beta = \gamma = 1/3$. We have thus determined optimal 'conditioning variables', i.e. the arguments $\bar{\mathbf{y}}$ and $\hat{\mathbf{w}}$ of the basis function $c_\alpha(\bar{\mathbf{y}}, \hat{\mathbf{w}})$, that are constructed in such a way that any measurement error-induced cross-terms cancel.

8

Thus, an unbiased estimator for the force projections in the presence of measurement errors is

$$\hat{F}_{\mu\alpha}^{(\text{w.m.e.})} = \langle \hat{a}_{\mu}^{(\text{w.m.e.})} c_\alpha(\bar{\mathbf{y}}(t), \hat{\mathbf{w}}(t)) \rangle \tag{S48a}$$

$$- \frac{1}{2} \left\langle (\partial_{v_\nu} c_\alpha(\bar{\mathbf{y}}(t), \hat{\mathbf{w}}(t))) \widehat{\sigma^2}_{\mu\nu}^{(\text{w.m.e.})}(\bar{\mathbf{y}}(t), \hat{\mathbf{w}}(t)) \right\rangle + \mathcal{O}(\Delta t, \eta^2)$$

$$\bar{\mathbf{y}} = \frac{1}{3}(\mathbf{y}(t - \Delta t) + \mathbf{y}(t) + \mathbf{y}(t + \Delta t)) \tag{S48b}$$

$$\hat{\mathbf{w}} = \frac{\mathbf{y}(t + \Delta t) - \mathbf{y}(t - \Delta t)}{2\Delta t} \tag{S48c}$$

$$\hat{\mathbf{a}}^{(\text{w.m.e.})} = \frac{\mathbf{y}(t + \Delta t) - 2\mathbf{y}(t) + \mathbf{y}(t - \Delta t)}{\Delta t^2} \tag{S48d}$$

As before, to infer the force field, we have to first find an estimator for the noise term $\widehat{\sigma^2}_{\mu\nu}^{(\text{w.m.e.})}$ that is not biased due to measurement errors.

## 3.2 Noise term

To derive an unbiased estimator for the noise amplitude in the presence of measurement errors, we follow a very similar line of thought to the derivation of the measurement error-corrected estimator for the force field. Specifically, using the increments of the process, we derive an estimator constructed such that the bias-terms due to the measurement error $\eta(t)$ vanish. However, in contrast to the force estimator, we now consider an estimator constructed from four points around $t$, $\{t - \Delta t, t, t + 2\Delta t, t + 3\Delta t\}$. This gives us three increments, rather than two as before, to construct our estimator. This additional freedom is required to construct an estimator that is not spoilt by measurement errors.

As before, we first start by constructing increments of the form

$$\Delta y_\mu^{(n)} = y_\mu(t + n\Delta t) - y_\mu(t) \tag{S49}$$

but now with $n = \{1, 2, 3\}$. We will later transform our results to a notation centered around $t$. Similar to Eq. (S21), we expand these increments, now including the measurement error

$$\Delta y_\mu^{(n)} = \Delta x_\mu^{(n)} + \Delta \eta_\mu^{(n)}$$
$$= v_\mu I_0^{(n)} + \sigma_{\mu\nu} I_{0\nu}^{(n)} + \Delta \eta_\mu^{(n)} + F_\mu I_{00}^{(n)} + (\partial_{v_\rho} \sigma_{\mu\nu}) \sigma_{\rho\tau} I_{0\nu\tau}^{(n)} + \mathcal{O}(\Delta t^{5/2}) \tag{S50}$$

Since we are aiming to infer the term $\sigma_{\mu\nu} I_{0\nu}^{(n)}$, which has zero average, we need to consider products of the increments (similar to the noise-free version (S31), where $\widehat{\sigma^2} \sim \hat{a}^2$).

$$\Delta_{\mu\nu}^{(n,m)} := \Delta y_\mu^{(n)} \Delta y_\nu^{(m)} \tag{S51}$$

We thus aim to construct an estimator of the form

$$\Delta t^{-3} \left\langle c_\alpha(\tilde{\mathbf{y}}, \check{\mathbf{w}}) \sum_{1 \leq m \leq n \leq 3} k_{mn} \Delta_{\mu\nu}^{(n,m)} \right\rangle \overset{!}{=} \sigma_{\mu\nu\alpha}^2 + \mathcal{O}(\Delta t, \eta^2) \tag{S52}$$

We therefore need to find the coefficients $k_{mn}$ for the linear combination of increment products and conditioning coordinates $\tilde{\mathbf{y}}$ and $\check{\mathbf{w}}$ such that all dynamical and measurement error cross-terms except for $\sigma^2$ cancel out to first order.

9

We start by expanding the increment products

$$
\begin{aligned}
\Delta_{\mu\nu}^{(n,m)} &= v_\mu v_\nu (nm\Delta t^2) + \sigma_{\mu\rho}\sigma_{\nu\tau} I_{0\rho}^{(n)} I_{0\tau}^{(m)} + \Delta\eta_\mu^{(n)} \Delta\eta_\nu^{(m)} + (v_\mu F_\nu mn^2 + v_\nu F_\mu m^2 n)\Delta t^3 \\
&\quad + (nv_\mu \sigma_{\nu\rho} I_{0\rho}^{(m)} + mv_\nu \sigma_{\mu\rho} I_{0\rho}^{(n)})\Delta t + (mv_\mu \Delta\eta_\nu^{(n)} + nv_\nu \Delta\eta_\mu^{(m)})\Delta t + \mathcal{O}(\Delta t^{7/2})
\end{aligned}
\tag{S53}
$$

Note that the last two terms in this expansion are zero on average, so one might think that we do not have to include them in the derivation of $k_{mn}$. This is correct in the case of constant noise. However, in the case of multiplicative noise, these terms correlate with terms in the expansion of the basis function $c_\alpha(\tilde{\mathbf{y}}, \check{\mathbf{w}})$, so we have to consider them.

Deriving the coefficients $k_{mn}$ is essentially a linear algebra problem, so we define the vectors

$$
\mathbf{d}_{\mu\nu}^{nm} = \left(\Delta_{\mu\nu}^{(1,1)}, \Delta_{\mu\nu}^{(2,2)}, \Delta_{\mu\nu}^{(3,3)}, \Delta_{\mu\nu}^{(1,3)}, \Delta_{\mu\nu}^{(2,3)}, \Delta_{\mu\nu}^{(1,2)}\right)^T
\tag{S54}
$$

$$
\mathbf{k}^{nm} = (k_{11}, k_{22}, k_{33}, k_{13}, k_{23}, k_{12})^T
\tag{S55}
$$

$$
\mathbf{t}_{\mu\nu} = \left(v_\mu v_\nu \Delta t^2, \sigma_{\mu\nu}^2 \Delta t^3, \Lambda\delta_{\mu\nu}, F_\mu v_\nu \Delta t^3, v_\mu \sigma_{\nu\rho} I_{0\rho}^{(1)}, v_\mu \sigma_{\nu\rho} I_{0\rho}^{(2)}, v_\mu \sigma_{\nu\rho} I_{0\rho}^{(3)}\right)^T
\tag{S56}
$$

Note, that in the definition of $\mathbf{t}_{\mu\nu}$ we have temporarily discarded the symmetry under exchange of $\mu, \nu$ for simplicity. We will later symmetrize our results to regain this symmetry. Furthermore, we have discarded the last term in Eq. (S53) in our definition of $\mathbf{t}_{\mu\nu}$. We will ignore this term in our derivation of $k_{mn}$ as we can take care of it through our choice of conditioning coordinates $\tilde{\mathbf{y}}$ and $\check{\mathbf{w}}$.

With these definitions, we explicitly evaluate the increment products:

$$
\mathbf{d}_{\mu\nu}^{nm} = \underline{\underline{R}}^T \cdot \mathbf{t}_{\mu\nu} + \mathcal{O}(\Delta t^{7/2})
\tag{S57}
$$

where

$$
\underline{\underline{R}} = \begin{pmatrix}
1 & 4 & 9 & 3 & 6 & 2 \\
1/3 & 8/3 & 9 & 4/3 & 14/3 & 5/6 \\
2 & 2 & 2 & 1 & 1 & 1 \\
2 & 16 & 54 & 12 & 30 & 6 \\
2 & 0 & 0 & 3 & 0 & 2 \\
0 & 4 & 0 & 0 & 3 & 1 \\
0 & 0 & 6 & 1 & 2 & 0
\end{pmatrix}
\tag{S58}
$$

Thus, a general estimator for the variable $V$ is given by solving the equation

$$
\hat{V}_{\mu\nu} = \mathbf{k}^{nm} \cdot \mathbf{d}_{\mu\nu}^{nm} \stackrel{!}{=} \boldsymbol{\ell}_V \cdot \mathbf{t}_{\mu\nu}
\tag{S59}
$$

for $\mathbf{k}^{nm}$. In our case the two quantities of interest are $\sigma^2$ and $\Lambda$, as we may also wish to infer the amplitude of the measurement error from the data. The constraint vectors $\boldsymbol{\ell}_V$ for these quantities are given by

$$
\boldsymbol{\ell}_{\sigma^2} = (0, 1, 0, 0, 0, 0)^T
\tag{S60}
$$

$$
\boldsymbol{\ell}_\Lambda = (0, 0, 1, 0, 0, 0)^T
\tag{S61}
$$

So far, we have derived everything in "(nm)-space", for increments as defined in Eq. (S49), which has the key advantage that they are easy to expand. However, for the final form of our

estimators, we choose a more natural definition of the increments,

$$\begin{aligned}
\Delta y_\mu^{(-)} &= y_\mu(t + \Delta t) - y_\mu(t) \\
\Delta y_\mu^{(0)} &= y_\mu(t + 2\Delta t) - y_\mu(t + \Delta t) \\
\Delta y_\mu^{(+)} &= y_\mu(t + 3\Delta t) - y_\mu(t + 2\Delta t)
\end{aligned} \tag{S62}$$

For this "$(+-)$-space", we define, similarly to before,

$$\mathbf{d}_{\mu\nu}^{+-} = \left(\Delta_{\mu\nu}^{(0,0)}, \Delta_{\mu\nu}^{(-,-)}, \Delta_{\mu\nu}^{(+,+)}, \Delta_{\mu\nu}^{(+,-)}, \Delta_{\mu\nu}^{(0,+)}, \Delta_{\mu\nu}^{(0,-)}\right)^T \tag{S63}$$

$$\mathbf{k}^{+-} = (k_{00}, k_{--}, k_{++}, k_{+-}, k_{0+}, k_{0-})^T \tag{S64}$$

We can transform between the two spaces using

$$\mathbf{d}_{\mu\nu}^{nm} = \underline{\underline{M}}\mathbf{d}_{\mu\nu}^{+-} \tag{S65}$$

with the transformation matrix

$$\underline{\underline{M}} = \begin{pmatrix} 0 & 1 & 0 & 0 & 0 & 0 \\ 1 & 1 & 0 & 0 & 0 & 2 \\ 1 & 1 & 1 & 2 & 2 & 2 \\ 0 & 1 & 0 & 1 & 0 & 1 \\ 1 & 1 & 0 & 1 & 1 & 2 \\ 0 & 1 & 0 & 0 & 0 & 1 \end{pmatrix} \tag{S66}$$

Thus, we need to solve the transformed equation

$$\underline{\underline{Q}}\mathbf{k}^{+-} = \ell_V \tag{S67}$$

where $\underline{\underline{Q}} = \underline{\underline{R}}(\underline{\underline{M}}^T)^{-1}$. Finally, we add two additional constraints to the matrix $\underline{\underline{Q}}$ which ensure that the final estimator is symmetric in the increments,

$$\underline{\underline{Q}}_{\text{sym}}\mathbf{k}_{\text{sym}}^{+-} = \ell_V^{\text{sym}} \tag{S68}$$

We can now solve for the coefficients:

$$\mathbf{k}_{\text{sym}}^{+-} = \underline{\underline{Q}}_{\text{sym}}^{\dagger}\ell_V^{\text{sym}} \tag{S69}$$

where $\underline{\underline{Q}}_{\text{sym}}^{\dagger}$ is the Moore-Penrose pseudoinverse of the non-square matrix $\underline{\underline{Q}}_{\text{sym}}$. This yields

$$\mathbf{k}_{\text{sym}}^{+-}(\sigma^2) = \frac{6}{11}(-1, 1, 1, -6, 1, 1)^T \tag{S70}$$

$$\mathbf{k}_{\text{sym}}^{+-}(\Lambda) = \frac{1}{44}(10, 1, 1, 8, -10, -10)^T \tag{S71}$$

With this solution for the coefficients, the estimator for $\sigma^2$ is now operational for the case of constant noise. The estimator for $\Lambda$ is valid equally in the case of multiplicative noise.

As we noted before, in the case of multiplicative noise, we need to adjust the conditioning variables $\tilde{\mathbf{y}}$ and $\check{\mathbf{w}}$ in order to avoid divergent biases due to the last term in Eq. (S53), similar

to the case of the force field. As our estimator is a four-point construct, we also construct the conditioning variables from four points:

$$\tilde{y}_\mu = \sum_{n=0}^{3} a_n y_\mu(t + n\Delta t) = \tilde{x}_\mu + \tilde{\eta}_\mu(\{a_n\}) \tag{S72}$$

$$\check{w}_\mu = \frac{1}{\Delta t}\left[b_1 \Delta y_\mu^{(-)} + b_2 \Delta y_\mu^{(0)} + b_3 \Delta y_\mu^{(+)}\right] = \check{v}_\mu + \frac{g_\mu^{(v)}(\eta, \{b_n\})}{\Delta t} \tag{S73}$$

where $\sum_{n=0}^{3} a_n = \sum_{n=1}^{3} b_n = 1$. Similarly to before (Eq. (S43)), we expand the basis functions

$$c_\alpha(\tilde{\mathbf{y}}, \check{\mathbf{w}}) = c_\alpha(\tilde{\mathbf{x}}, \check{\mathbf{v}}) + (\partial_{x_\nu} c_\alpha)\tilde{\eta}_\nu(\{a_n\}) + (\partial_{v_\nu} c_\alpha)\frac{g_\nu^{(v)}(\eta, \{b_n\})}{\Delta t} + \mathcal{O}(\eta^2) \tag{S74}$$

The remaining bias in our estimator (Eq. (S52)) is due to the last term in Eq. (S53),

$$q_{\mu\nu}^{(m,n)} = (mv_\mu \Delta \eta_\nu^{(n)} + nv_\nu \Delta \eta_\mu^{(m)})\Delta t \tag{S75}$$

We define

$$\mathbf{q}_{\mu\nu}^{nm} = \left(q_{\mu\nu}^{(1,1)}, q_{\mu\nu}^{(2,2)}, q_{\mu\nu}^{(3,3)}, q_{\mu\nu}^{(1,3)}, q_{\mu\nu}^{(2,3)}, q_{\mu\nu}^{(1,2)}\right)^T \tag{S76}$$

and can thus write

$$\Delta t^{-3}\left\langle c_\alpha(\tilde{\mathbf{y}}, \check{\mathbf{w}})\mathbf{k}_{\text{sym}}^{mn} \cdot \mathbf{d}_{\mu\nu}^{nm}\right\rangle = \langle \sigma_{\mu\nu}^2 c_\alpha(\tilde{\mathbf{x}}, \check{\mathbf{v}})\rangle$$

$$+ \Delta t^{-3}\left\langle \mathbf{k}_{\text{sym}}^{mn} \cdot \mathbf{q}_{\mu\nu}^{nm}\left((\partial_{x_\rho} c_\alpha)\tilde{\eta}_\rho(\{a_n\}) + (\partial_{v_\rho} c_\alpha)\frac{g_\rho^{(v)}(\eta, \{b_n\})}{\Delta t}\right)\right\rangle + \mathcal{O}(\Delta t, \eta^2) \tag{S77}$$

This shows that the bias terms are of order $\Delta t^{-3}$ and $\Delta t^{-4}$, and thus diverge in the limit $\Delta t \to 0$. We now need to find coefficients $\{a_n\}$ and $\{b_n\}$ such that

$$\left\langle \mathbf{k}_{\text{sym}}^{mn} \cdot \mathbf{q}_{\mu\nu}^{nm} \tilde{\eta}_\rho(\{a_n\})\right\rangle = 0 \tag{S78}$$

$$\left\langle \mathbf{k}_{\text{sym}}^{mn} \cdot \mathbf{q}_{\mu\nu}^{nm} \frac{g_\rho^{(v)}(\eta, \{b_n\})}{\Delta t}\right\rangle = 0 \tag{S79}$$

We start by explicitly evaluating $\mathbf{q}_{\mu\nu}^{nm}$:

$$\mathbf{q}_{\mu\nu}^{nm} = \underline{\underline{E}} \cdot \mathbf{h}_\mu \cdot (v_\nu \Delta t) \tag{S80}$$

where

$$\underline{\underline{E}} = \begin{pmatrix} -2 & 2 & 0 & 0 \\ -4 & 0 & 4 & 0 \\ -6 & 0 & 0 & 6 \\ -5 & 0 & 3 & 2 \\ -4 & 3 & 0 & 1 \\ -3 & 2 & 1 & 0 \end{pmatrix} \tag{S81}$$

and

$$\mathbf{h}_\mu = \left(\eta_\mu(t), \eta_\mu(t + \Delta t), \eta_\mu(t + 2\Delta t), \eta_\mu(t + 3\Delta t)\right)^T. \tag{S82}$$

12

We first focus on the conditioning of the configurational (position-like) coordinate, *i.e.* solving Eq. (S78). Defining $\mathbf{a} = (a_0, a_1, a_2, a_3)^T$, Eq. (S78) becomes

$$\left\langle \left(\mathbf{k}_{\text{sym}}^{\text{mn}} \cdot \mathbf{q}_{\mu\nu}^{\text{nm}}\right) (\mathbf{a} \cdot \mathbf{h}_\rho) \right\rangle = 0 \tag{S83}$$

Evaluating

$$\mathbf{k}_{\text{sym}}^{\text{mn}} \cdot \mathbf{q}_{\mu\nu}^{\text{nm}} = \underline{\underline{E}}^T \cdot \underline{\underline{M}} \cdot \mathbf{k}_{\text{sym}}^{+-}(\sigma^2) \tag{S84}$$

$$= \frac{1}{11}(-30, 36, 42, -48)^T \tag{S85}$$

shows that Eq. (S78) is solved by

$$\mathbf{a} = \frac{1}{4}(1, 1, 1, 1)^T. \tag{S86}$$

Next, we find the conditioning of the velocity coordinate, *i.e.* solving Eq. (S79). Defining $\mathbf{b} = (b_1, b_2, b_3)^T$, Eq. (S79) becomes

$$\left\langle \left(\mathbf{k}_{\text{sym}}^{\text{mn}} \cdot \mathbf{q}_{\mu\nu}^{\text{nm}}\right) \frac{\mathbf{b} \cdot \underline{\underline{F}} \cdot \mathbf{h}_\rho}{\Delta t} \right\rangle = 0 \tag{S87}$$

where

$$\underline{\underline{F}} = \begin{pmatrix} -1 & 1 & 0 & 0 \\ 0 & -1 & 1 & 0 \\ 0 & 0 & -1 & 1 \end{pmatrix} \tag{S88}$$

are the coefficients of the measurement error $\mathbf{h}$ in the velocity estimator $\check{\mathbf{w}}$. Evaluating

$$\underline{\underline{F}} \cdot \mathbf{k}_{\text{sym}}^{\text{mn}} \cdot \mathbf{q}_{\mu\nu}^{\text{nm}} = \underline{\underline{F}} \cdot \underline{\underline{E}}^T \cdot \underline{\underline{M}} \cdot \mathbf{k}_{\text{sym}}^{+-}(\sigma^2) \tag{S89}$$

$$= \frac{6}{11}(11, 1, -15)^T \tag{S90}$$

shows that Eq. (S79) is solved by

$$\mathbf{b} = \frac{1}{6}(1, 4, 1)^T. \tag{S91}$$

Summarizing, an unbiased estimator for the projection coefficients of the multiplicative noise amplitude in the presence of measurement error is given by

$$\widehat{\sigma^2}_{\mu\nu\alpha}^{\text{(w.m.e.)}} = \Delta t^{-3} \left\langle c_\alpha(\tilde{\mathbf{y}}, \check{\mathbf{w}}) \mathbf{k}_{\text{sym}}^{+-} \cdot \mathbf{d}_{\mu\nu}^{+-} \right\rangle \tag{S92a}$$

$$\mathbf{k}_{\text{sym}}^{+-} = \frac{6}{11}(-1, 1, 1, -6, 1, 1)^T \tag{S92b}$$

$$\tilde{\mathbf{y}} = \frac{1}{4}(\mathbf{y}(t - \Delta t) + \mathbf{y}(t) + \mathbf{y}(t + \Delta t) + \mathbf{y}(t + 2\Delta t)) \tag{S92c}$$

$$\check{\mathbf{w}} = \frac{\Delta \mathbf{y}^{(-)} + 4\Delta \mathbf{y}^{(0)} + \Delta \mathbf{y}^{(+)}}{6\Delta t} \tag{S92d}$$

with $\Delta \mathbf{y}^{(+/0/-)}$ as defined by Eq. (S62) and

$$\mathbf{d}_{\mu\nu}^{+-} = \left(\Delta y_\mu^{(0)} \Delta y_\nu^{(0)}, \Delta y_\mu^{(-)} \Delta y_\nu^{(-)}, \Delta y_\mu^{(+)} \Delta y_\nu^{(+)}, \Delta y_\mu^{(+)} \Delta y_\nu^{(-)}, \Delta y_\mu^{(0)} \Delta y_\nu^{(+)}, \Delta y_\mu^{(0)} \Delta y_\nu^{(-)}\right)^T \tag{S93}$$

13

## 3.3 Scaling of the inference error with the measurement error amplitude

To determine the critical measurement error amplitude at which the estimators fail, we investigate the scaling of the error curves with the observation time interval $\Delta t$ for the damped harmonic oscillator. We find that the error curves of the estimator without noise correction (section 2.2) collapse with $\sigma \Delta t^{3/2}$, while the curves of the estimator with noise correction (section 3.1) collapse with $\Delta t$.

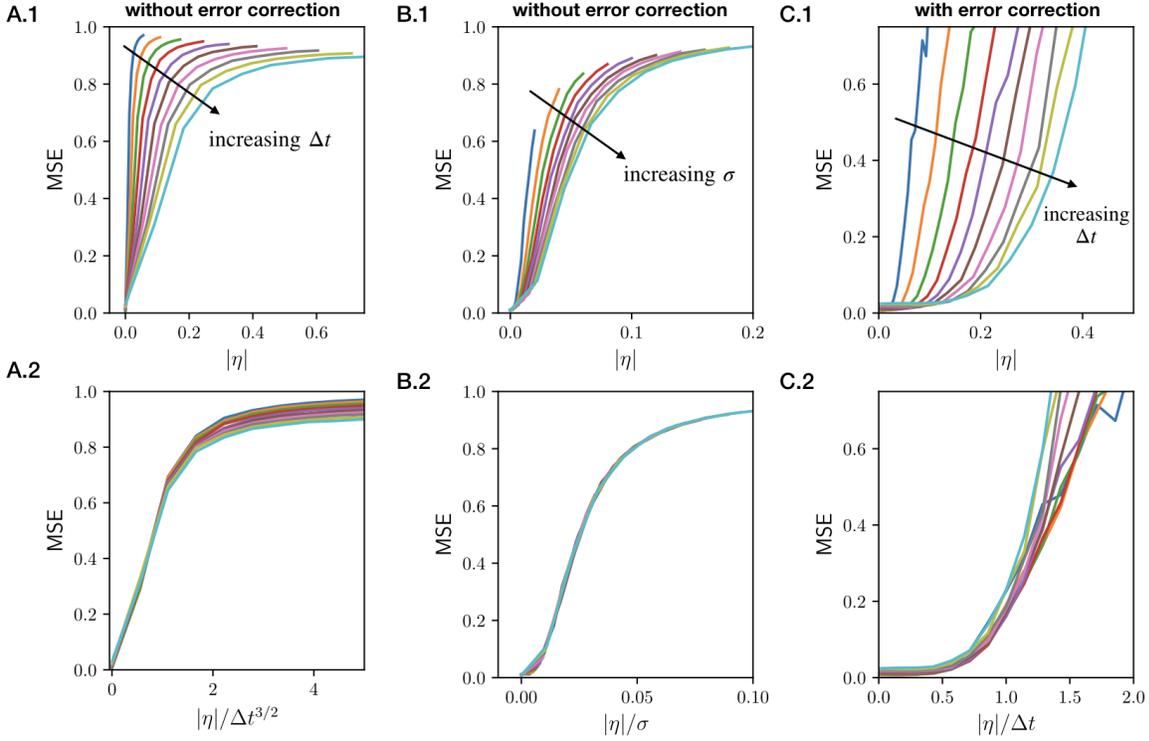

Figure S1: **Error scaling of the force estimator in the presence of measurement error.** **A.** *Top:* mean-square-error for the estimator without noise correction (section 2.2) as a function of the measurement error amplitude $|\eta|$ for different values of $\Delta t$. *Bottom:* Data collapse by dividing by $\Delta t^{3/2}$. **B.** *Top:* same plot as in B, but for different values of $\sigma$. *Bottom:* Data collapse by dividing by $\sigma$. **C.** *Top:* mean-square-error along the trajectory for the estimator with noise correction (section 3.1) as a function of the measurement error amplitude $|\eta|$ for different values of $\Delta t$. *Bottom:* Data collapse by dividing by $\Delta t$.

## 4 Choosing the basis size $n_b$

We perform inference by projecting the dynamics onto a finite set of $n_b$ basis functions $\{b_\alpha(\mathbf{x}, \mathbf{v})\}$ where the index $\alpha$ runs over all basis functions in the set, $1 \leqslant \alpha \leqslant n_b$. Thus, we need to choose a value for $n_b$, and this will clearly influence the accuracy of the inference, leading to the question: is there an optimal choice $n_b^{(\mathrm{opt})}$? A criterion to choose $n_b^{(\mathrm{opt})}$ was proposed for overdamped stochastic processes in ref. [1], which is based on the empirical estimate of the information content in the observed trajectory. Here, we show that this criterion similarly applies to underdamped stochastic systems.

The larger $n_b$, the more accurately it can capture the features of the force field. Thus, the information $\hat{I}_b = \frac{\tau}{2}\hat{\sigma}^{-2}_{\mu\nu}\hat{F}_{\mu\alpha}\hat{F}_{\nu\alpha}$ captured by the force field representation [1] increases with $n_b$. However, for a finite trajectory, the error in the inferred force field will also increase with $n_b$. Thus, we expect a trade-off between the inference error and the completeness of the force field projection. To choose $n_b^{\text{opt}}$, we therefore maximize the information $\hat{I}_b$ that can be statistically resolved, by determining the basis size which maximizes $\hat{I}_b - \delta\hat{I}_b$, where $\delta\hat{I}_b \approx \sqrt{2\hat{I}_b + N_b^2/4}$ is the typical error in the inferred information and $N_b = d \times n_b$ is the number of parameters to infer [1].

In Fig. S2, we plot the inference error $\delta\hat{F}^2/\hat{F}^2$ as a function of the number of parameters $N_b$ for the 1D Van der Pol oscillator projected on a basis consisting of Fourier components in $x$ and polynomials in $v$. For all trajectory lengths, we see the expected behaviour: at small $N_b$, the error first decreases with increasing $N_b$, since it is dominated by underfitting. Beyond the optimum, the error increases with $N_b$, due to the increasing inference error. Clearly, the optimal basis size $n_b^{(\text{opt})}$ increases with the length of the trajectory, as more information on the features of the force field becomes available. We find that the overfitting criterion to maximize $\hat{I}_b - \delta\hat{I}_b$ yields an accurate prediction of the optimal basis size (starred symbols in Fig. S2).

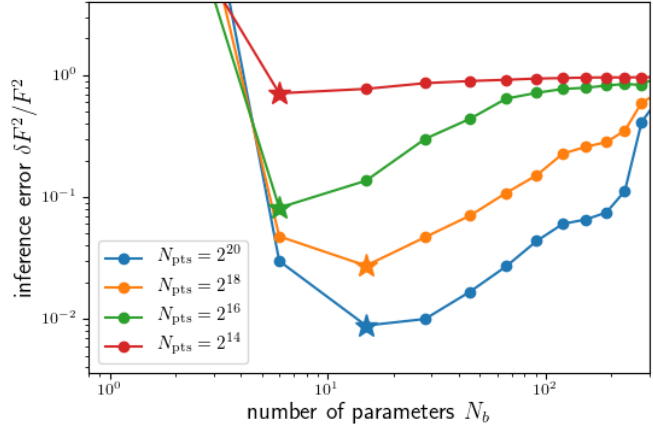

Figure S2: **Quantification of the inference performance as a function of basis size.** Here we study the 1D Van der Pol oscillator $\dot{v} = \mu(1 - x^2)v - x + \sigma\xi(t)$ as an example. The inference error $\delta\hat{F}^2/\hat{F}^2$ is plotted as a function of number of parameters $N_b$ used in the projection basis, consists of Fourier components in $x$ and polynomials in $v$. Each curve corresponds to the result obtained from a single trajectory with the number of time frames indicated in the legend. Starred symbols indicate the predicted optimal basis size $n_b^{(\text{opt})}$ determined by maximizing $\hat{I}_b - \delta\hat{I}_b$.

# 5 Inference from experimental single cell trajectories

Here, we discuss the inference from experimental single cell trajectories shown in Fig. 2 of the main text in more detail. Specifically, we show that the experimental trajectories contain enough information to perform Underdamped Langevin Inference, and that the inferred models can be inferred self-consistently. Importantly, here the term $F_\mu(\mathbf{x}, \mathbf{v})$ corresponds to the underlying deterministic dynamics of the system, and not to a physical force. We therefore call it the "deterministic term" of the dynamics. Details on cell culture, experimental protocols and tracking procedures can be found in ref. [5].

## 5.1 Information content of experimental single cell trajectories

As discussed in the main text, the observed trajectories are limited in length due to the finite life-time of a single cell, up until the point where it divides. The expected mean-squared-error in the inferred flow field projected onto a basis $b$ is given by

$$\delta \hat{F}^2 / \hat{F}^2 \sim N_b / 2\hat{I}_b, \tag{S94}$$

where $N_b$ is the number of degrees of freedom in the basis $b$, i.e. the number of fit parameters. $\hat{I}_b$ is the empirical estimate of the information content of the trajectory of length $\tau$, given by

$$\hat{I}_b = \frac{\tau}{2} \hat{\sigma}_{\mu\nu}^{-2} \hat{F}_{\mu\alpha} \hat{F}_{\nu\alpha}, \tag{S95}$$

measured in natural information units (1 nat $= 1/\log 2$ bits). We estimate this information by projecting onto a third-order polynomial basis, and find that the average information per trajectory is 94.2 nats (Fig. S3). To perform accurate inference, we need $\hat{I}_b \gg N_b$. In previous work [5, 8], we inferred models averaged over large numbers of cell trajectories using a basis of $30 \times 30$ coarse-grained bins, i.e. $N_b = 900$. Thus, single-cell inference was not possible with this approach. In contrast, here we use the partial information to guide a principled selection of basis functions, which shows that most of the information is captured by a symmetrised third-order polynomial basis $\{x, v, x^3, x^2v, xv^2, v^3\}$. Thus, we infer $N_b = 6$ parameters and the criterion $\hat{I}_b \gg N_b$ is fulfilled.

## 5.2 Self-consistency test of the single-cell inference

To test whether the inferred single-cell models are self-consistent, we simulate trajectories based on the inferred dynamics (Fig. S4). These trajectories perform stochastic transitions on a similar time-scale to the experimental trajectories and exhibit similar oscillation loops in the $xv$-phase space (Fig. S4E,F). To test model stability, we simulate trajectories of the same length as the experimental ones and sample at the same time interval as in experiment ($\Delta t = 10$ min). From these trajectories, we then infer a bootstrapped flow field, which exhibits similar qualitative features as the original flow field inferred from experiments (Fig. S4G). To quantify this, we directly compare the values of the bootstrapped $F(x, v)$ relative to the experimentally inferred $F(x, v)$ along the experimental trajectory (Fig. S4H), which shows strong correlation with a typical mean-squared-error of order 0.3. Thus, ULI with a symmetrised third-order polynomial basis provides robust, self-consistent models for single-cell trajectories.

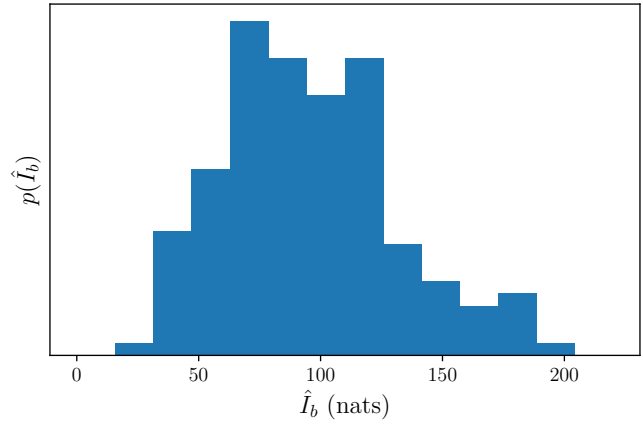

Figure S3: **Information content of single cell trajectories.** Histogram of the information content $\hat{I}_b$ of $N = 149$ single cell trajectories, obtained by projecting onto a third-order polynomial basis. The information is measured in natural information units (1 nat = $1/\log 2$ bits). The average information per trajectory is 94.2 nats.

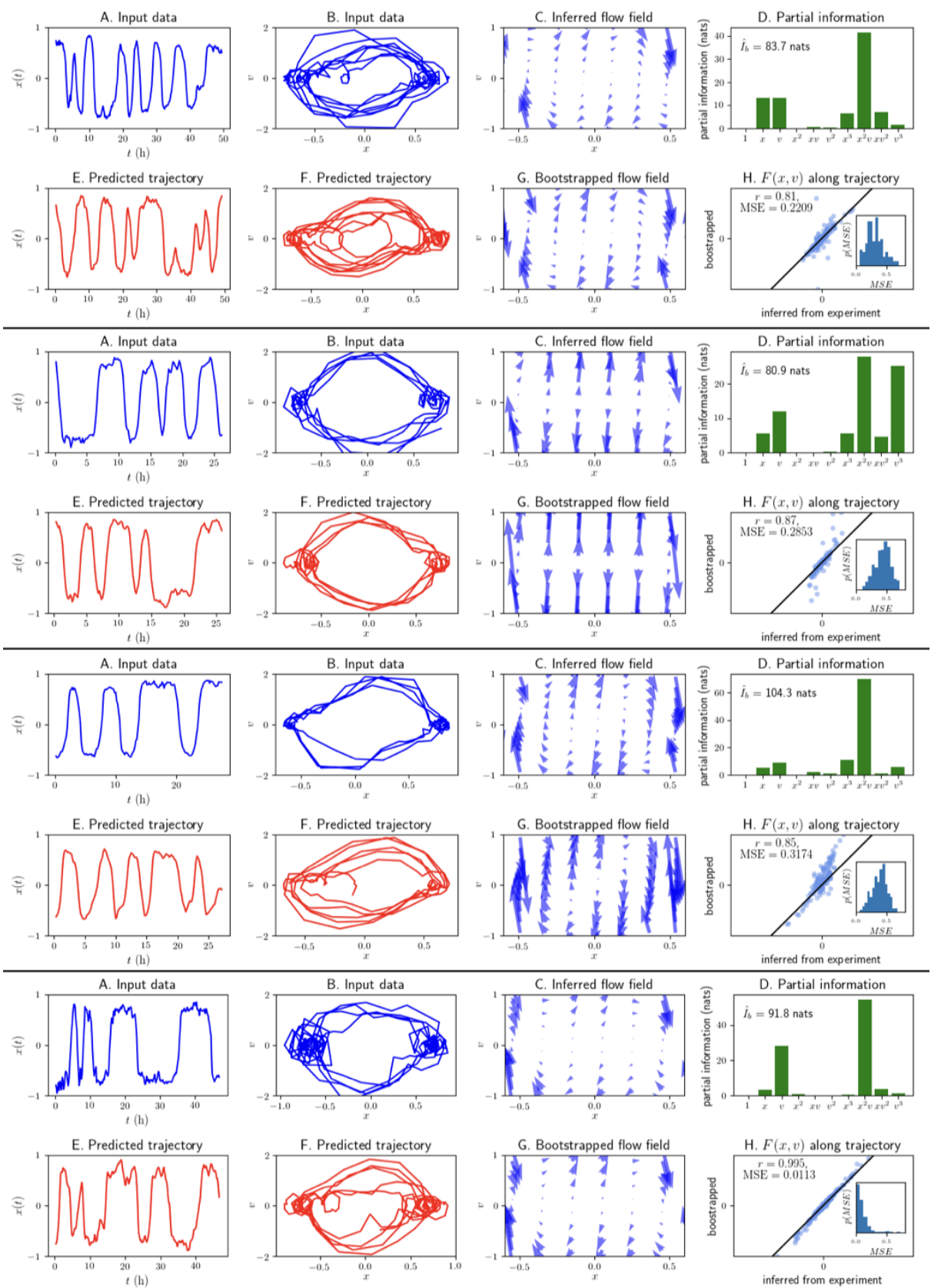

Figure S4: **Inferring single-cell models from two-state migration trajectories. A.** Experimentally recorded trajectory of the cell nucleus position, sampled at a time-interval $\Delta t = 10$ min. **B.** $xv$-plot of the trajectory shown in A. **C.** Flow field inferred from the trajectory in A using ULI with a symmetrised third-order polynomial basis, $\{x, v, x^3, x^2v, xv^2, v^3\}$. **D.** Partial information of the trajectory shown in A, projected onto a third-order polynomial basis. The total estimated information $\hat{I}_b$ of the trajectory is given. **E.** Trajectory simulated using the inferred model, consisting of the deterministic flow field in C and the inferred constant noise amplitude. The process is simulated at a small time-interval and subsequently sampled at the experimental time-interval $\Delta t = 10$ min. **F.** $xv$-plot of the simulated trajectory shown in E. **G.** Bootstrapped flow field inferred from the simulated trajectory in E using ULI with a symmetrised third-order polynomial basis. **H.** Scatter plot of the deterministic term evaluated at the points visited by the experimental trajectory, comparing the flow field inferred from experiment against the bootstrapped result. The mean-squared-error (*MSE*) and the Pearson *r*-coefficient are given. *Inset:* histogram of the mean-squared-error of $N = 300$ bootstrap realizations. The four subfigures correspond to four individual cell trajectories. The top subfigure corresponds to the trajectory shown in Fig. 2 of the main text.

## 6 Model details and simulation parameters for numerical results

To benchmark ULI, we apply it to several canonical examples of underdamped stochastic processes (Fig. 1-3). To simulate these processes, we employ a simple discretization scheme

$$\mathbf{x}(t + \mathrm{d}t) = \mathbf{x}(t) + \mathbf{v}(t)\mathrm{d}t \tag{S96}$$

$$\mathbf{v}(t + \mathrm{d}t) = \mathbf{v}(t) + \mathbf{F}(\mathbf{x}(t), \mathbf{v}(t))\mathrm{d}t + \sqrt{\mathrm{d}t}\,\underline{\sigma}(\mathbf{x}(t), \mathbf{v}(t)) \cdot \boldsymbol{\zeta}(t) \tag{S97}$$

where $\boldsymbol{\zeta}$ is a vector of independent Gaussian random numbers with zero mean and unit variance. We simulate this equation with a small time interval $\mathrm{d}t$ to ensure numerical stability. To generate a realistic experimental position trajectory, we sample the simulated trajectory with a larger interval $\Delta t$ and add an uncorrelated measurement error to the positions. We use $\mathrm{d}t = \Delta t/20$ throughout. Thus, ULI only has access to the trajectory

$$\{\mathbf{y}(0), \mathbf{y}(\Delta t), \mathbf{y}(2\Delta t), ..., \mathbf{y}(\tau - \Delta t), \mathbf{y}(\tau)\} \quad \text{where} \quad \mathbf{y}(t) = \mathbf{x}(t) + \boldsymbol{\eta}(t) \tag{S98}$$

and $\boldsymbol{\eta}$ is a vector of independent Gaussian random numbers with zero mean and unit variance, such that

$$\eta_\mu(t)\eta_\nu(t') = \Lambda \delta_{\mu\nu} \delta(t - t') \tag{S99}$$

and we define $|\eta| = \sqrt{\Lambda}$. The total duration of a trajectory with $N_{\text{steps}}$ observation points given by $\tau = N_{\text{steps}}\Delta t$.

### 6.1 Damped harmonic oscillator (Fig. 1)

We simulate the 1D stochastic damped harmonic oscillator,

$$\dot{v} = -\gamma v - kx + \sigma \xi \tag{S100}$$

We use $\gamma = k = \sigma = 1$ in all panels. Furthermore, we use
Fig. 1A-C: $N_{\text{steps}} = 10^3, \Delta t = 0.1, |\eta| = 0$

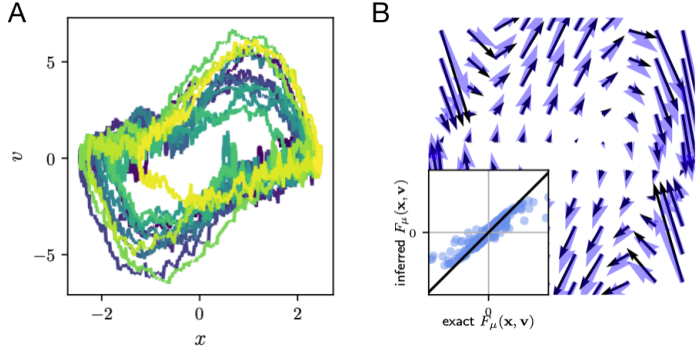

Figure S5: **Inferring Van der Pol dynamics with a Fourier basis. A.** Same trajectory as in Fig. 2A in the main text. **B.** ULI applied to the trajectory in A with basis functions $b = \{\sin(a_x x), \sin(a_v v), \sin(a_x x)\cos(a_v v), \cos(a_x x)\sin(a_v v)\}$. In general, a reasonable choice of the non-linear parameters $a_x, a_v$ is $a_x = 2\pi/L_x, a_v = 2\pi/L_v$, where $L_x, L_v$ are the widths of the sampled phase space in the $x$ and $v$ directions, respectively. From the trajectory in A, we see that $L_x = 6, L_v = 12$ are reasonable choices. *Inset:* inferred components of the force along the trajectory *versus* the exact values.

Fig. 1F-H: $N_{\text{steps}} = 10^3, \Delta t = 0.1, |\eta| = 0.02$
Fig. 1D: $\Delta t = 0.1, |\eta| = 0$
Fig. 1E: $\tau = 10^2, |\eta| = 0$
Fig. 1J: $\Delta t = 0.1, |\eta| = 0.02$
Fig. 1K: $\Delta t = 0.1, N_{\text{steps}} = 10^4$

## 6.2 Van der Pol oscillator (Fig. 2)

For the Van der Pol oscillator, we use $\kappa = 2, \sigma = 1$ throughout.
In Fig. 2A,B, we simulate the 1D Van der Pol oscillator

$$\dot{v} = \kappa(1 - x^2)v - x + \sigma\xi \tag{S101}$$

with $\Delta t = 0.01, N_{\text{steps}} = 10^4, |\eta| = 0.002$. As shown in Supplementary Fig. S5, we recover the dynamics similarly well using a Fourier basis rather than a polynomial basis in the inference.

In Fig. 2C, we simulate the $d$-dimensional Van der Pol oscillator $F_\mu(\mathbf{x}, \mathbf{v}) = \kappa_\mu(1 - x_\mu^2)v_\mu - x_\mu$ (no summation over $\mu$, $1 \leqslant \mu \leqslant d$) with $d = 1...6$. Here, we use the same parameters as for the 1D Van der Pol oscillator, and take $\kappa_\mu = 2 \,\forall\, \mu$.

In Fig. 2 G-J, we simulate the 1D Van der Pol oscillator with multiplicative noise

$$\dot{v} = \mu(1 - x^2)v - x + \sigma(x, v)\xi \tag{S102}$$

where $\sigma^2(x, v) = \sigma_0 + \sigma_x x^2 + \sigma_v v^2$. We use $\sigma_0 = 1, \sigma_x = 0.3, \sigma_v = 0.1, \Delta t = 0.01, N_{\text{steps}} = 10^4, |\eta| = 0.002$.

## 6.3 Interacting flocks (Fig. 3)

The model we simulate is a three-dimensional flock of $N = 27$ aligning self-propelled particles, with "soft Lennard-Jones"-type interactions. The particles are initialized on a $3 \times 3 \times 3$ grid

20

with zero velocity. The force on particle *i* is given by

$$\mathbf{F}_i = \gamma(v_0^2 - |\mathbf{v}_i|^2)\mathbf{v}_i + \sum_{j \neq i} \left[ \epsilon_0 \frac{1 - (r/r_0)^3}{(r_{ij}/r_0)^6 + 1} \mathbf{r}_{ij} + \epsilon_1 \exp(-r_{ij}/r_1)\mathbf{v}_{ij} \right] \quad (S103)$$

where $\mathbf{r}_{ij} = \mathbf{r}_j - \mathbf{r}_i$, $\mathbf{v}_{ij} = \mathbf{v}_j - \mathbf{v}_i$, while the noise $\sigma \xi_i(t)$ on each particle is isotropic and uncorrelated with others. We choose the parameters $\gamma = 1$, $v_0 = 1.5$, $\epsilon_0 = 4$, $r_0 = 2$, $\epsilon_1 = 1$, $r_1 = 3$ and $\sigma = 1$, which result in a flocking behavior similar to that of bird flocks. The simulation is performed with a time step $dt = 0.005$. It is run for 2000 steps to reach steady state before recording, then the trajectory consisting in 1000 time points with time interval $\Delta t = 0.02$ is recorded.

For the inference, we employ a translation-invariant basis with single-particle and pair interaction terms that is invariant under particle exchange $i \leftrightarrow j$, such that

$$F_{i,\mu} \approx F_{\mu\alpha}^{(1)} c_\alpha^{(1)}(\mathbf{v}_i) + F_{\mu\alpha}^{(2)} \sum_{j \neq i} c_\alpha^{(2)}(\mathbf{x}_i - \mathbf{x}_j, \mathbf{v_i}, \mathbf{v_j}) \quad (S104)$$

The single particle fitting functions are chosen to be polynomials of order up to 3 in the velocity (20 functions). The pair interactions are chosen to be of two kinds: radial functions $\sum_j k(r_{ij})\mathbf{r}_{ij}$ and velocity alignment functions $\sum_j k(r_{ij})\mathbf{v}_{ij}$. We choose the same set of fitting kernels $k(r)$ for both radial force and alignment, $k_n(r) = \exp(-r/r_n)$ with $r_n = 0.5n$ and $n = 1 \ldots 8$. The outcome of force inference is not very sensitive to this choice; $r$-dependent Gaussian kernels centered at different radii gives similar results. These result in 8 functions for each component of the vectors $\mathbf{r}_{ij}$ and $\mathbf{v}_{ij}$, hence 48 functions pair interaction functions. There are thus 68 functions in the basis, and thus 204 fit parameters for the force field. Inferring the noise tensor and these fit coefficients, we find that the total information in the trajectory presented in Fig. 3 of the main text is $\hat{I} = 320,000$ nats – more than enough to precisely resolve these parameters. Indeed, we find a mean-squared error on the force of 0.015 along the trajectory; this error could be reduced by adding more functions to the basis, or by using longer trajectories, as shown in the convergence plot in Fig. 3E.



\end{document}